\documentclass[12pt,twoside]{article}
\usepackage{style} 
\newtheorem{theorem}{Theorem}[section]

\def\qed{\rule{2mm}{2mm}}

\begin{document}
%%%%%%%%%%%%%%%%%%%%%%%%%%%%%%%%%%%%%%%%%%%%%%%%%%%%%%%%%%%%%%%%%
% TITLE PAGE
%%%%%%%%%%%%%%%%%%%%%%%%%%%%%%%%%%%%%%%%%%%%%%%%%%%%%%%%%%%%%%%%%
\begin{titlepage}

\begin{spacing}{0.9}
\title{\textbf{Confidence Intervals for Seroprevalence
	\thanks{We acknowledge funding from the National Science Foundation under the Graduate Research Fellowship Program and under the grants MMS-1949845 and SES-1530661.}}}

\runningheads{Confidence Intervals for Seroprevalence}{DiCiccio, Ritzwoller, Romano, and Shaikh}

\author{
\begin{tabular}[t]{c@{\extracolsep{2em}}c} 
\large 
Thomas J. DiCiccio &  David M. Ritzwoller \\ \vspace{-0.7em}
\small{\it Department of Social Statistics} & \small{\it Graduate School of Business} \\ \vspace{-0.7em}
\small{\it Cornell University} & \small{\it Stanford University} \\
\small{tjd9@cornell.edu} & \small{ritzwoll@stanford.edu}\\
Joseph P. Romano &  Azeem M. Shaikh \\ \vspace{-0.7em}
\small{\it Departments of Statistics and Economics} & \small{\it Department of Economics} \\ \vspace{-0.7em}
\small{\it Stanford University} & \small{\it University of Chicago} \\
\small{romano@stanford.edu} & \small{amshaikh@uchicago.edu}
\end{tabular}
}

\date{\small \today}
\maketitle
\setcounter{page}{0}
\thispagestyle{empty}

\vspace{-0.3in}
\begin{abstract}
This paper concerns the construction of confidence intervals in standard seroprevalence surveys. In particular, we discuss methods for constructing confidence intervals for the proportion of individuals in a population  infected with a disease using a sample of antibody test results and measurements of the test's false positive and false negative rates. We begin by documenting erratic behavior in the coverage probabilities of standard Wald and percentile bootstrap intervals when applied to this problem. We then consider two alternative sets of intervals constructed with test inversion. The first set of intervals are approximate, using either asymptotic or bootstrap approximation to the finite-sample distribution of a chosen test statistic. We consider several choices of test statistic, including maximum likelihood estimators and generalized likelihood ratio statistics. We show with simulation that,  at empirically relevant parameter values and sample sizes, the coverage probabilities for these intervals are close to their nominal level and are approximately equi-tailed.  The second set of intervals are shown to contain the true parameter value with probability at least equal to the nominal level, but can be conservative in finite samples.

\end{abstract}
\end{spacing}
\end{titlepage}

%%%%%%%%%%%%%%%%%%%%%%%%%%%%%%%%%%%%%%%%%%%%%%%%%%%%%%%%%%%
\section{Introduction}\label{section:intro} 
%%%%%%%%%%%%%%%%%%%%%%%%%%%%%%%%%%%%%%%%%%%%%%%%%%%%%%%%%%%

Effective public health policy requires accurate measurement of the spread of infections diseases \citep{fauci2020covid, peeling2020serology}. Seroprevalence surveys, in which antibody tests are administered to samples of individuals from populations of interest, are a practical and widely applied strategy for assessing the progression of a pandemic \citep{krammer2020serology, alter2020power}. However, antibody tests, which detect the presence of viral antibodies in blood samples, are imperfect.\footnote{An early systematic review of the accuracy of SARS-CoV-2 antibody tests is given in \cite{deeks2020antibody}, highlighting several methodological limitations. False positive and negative rates for five leading SARS-CoV-2 immunoassays are measured in \cite{ainsworth2020performance}. Estimates of false positive rates ranged from $0.1\%$ to $1.1\%$. Estimates of false negative rates ranged from $0.9\%$ to $7.3\%$.} Accounting for the variation in the results of seroprevalence surveys induced by this imperfection is important for informative assessment of the uncertainty in measurements of the spread of infectious diseases. 

In this paper, we study the construction of confidence intervals in standard seroprevalence surveys. Given the public interest in explicit representations of disease incidence, our objective is to analyze the accuracy of various methods of constructing confidence intervals, so that results from empirical analyses can be reported with statistical precision. We demonstrate that some methods based on test inversion offer advantages relative to more standard confidence interval constructions in terms of the accuracy and validity of their coverage probabilities. 

In a standard seroprevalence survey, the proportion of a population that has been infected with a disease is a smooth function of the parameters of three independent binomial trials. Although it may be expected that standard approaches to confidence interval construction are well suited for such a simple parametric problem, in Section \ref{sec: problem} we demonstrate with simulation that standard Wald and percentile bootstrap confidence intervals have erratic coverage probabilities at empirically relevant parameter values  and sample sizes when applied to this problem. 

In fact, as documented in \cite{brown2001interval}, erratic coverage probabilities for  confidence intervals constructed using standard methods surface even in the context of inference on a single binomial parameter. Bootstrap (and other) methods that are typically second-order correct in continuous problems may not achieve this accuracy in discrete problems.\footnote{Usually, claims of second-order correctness are made based on Edgeworth expansions \citep{hall2013bootstrap}. However, in discrete settings, Cram\'er's condition, a necessary condition for the application of an Edgeworth expansion, fails, and second-order accuracy may not be achievable. For example, atoms in the binomial distribution based on $n$ trials have order $n^{-1/2}$,  so expansions to order $n^{-1}$ must account for this discreteness.} Additionally, when  a binomial random variable has parameter value near zero or one, even first-order approximations to its limiting distribution are not normal, while many standard methods for constructing confidence intervals rely, either explicitly or implicitly, on a normal approximation holding. As the parameter of interest in a seroprevalence survey is a function of three binomial parameters, inference in this setting is more challenging than for a single binomial proportion.

To address the erratic coverage probabilities in standard confidence interval constructions, we consider several alternative approaches based on test inversion. A test inversion confidence interval for a parameter $\theta$ consists of the set of points $\theta_0$ for which the null hypothesis $H( \theta_0 ):\theta = \theta_0$ is not rejected. For parameters $\theta$ where the corresponding null hypothesis $H( \theta_0 )$ is simple, the application of test inversion is  straightforward. However, when the corresponding null hypothesis is composite, as in the case of seroprevalence, the application of test inversion is not immediate.

Thus, in Section \ref{sec: inversion intervals} we explore the general problem of test inversion for parameters whose corresponding null hypotheses are composite.  We consider both methods based on asymptotic or bootstrap approximation and methods with finite-sample guarantees. In the later case, a maximization of $p$-values over a nuisance parameter space is required, as in \cite{bergerboos:1994} and \cite{silvapulle:1996}. In practice, this maximization is carried out over a discrete grid. We provide a refinement to such an approximation that maintains the finite-sample coverage requirement. We take particular care in requiring that the confidence intervals that we develop behave well at both endpoints; that is, we require that they are equi-tailed.\footnote{A $1-\alpha$ confidence interval is equi-tailed if the probabilities that the parameter exceeds the upper endpoint or is below the lower endpoint of the interval are both near or below $\alpha/2$. That is, an equi-tailed $1-\alpha$ confidence interval should be given by the set of points satisfying both an upper and a lower confidence bound, each at level $1-\alpha/2$.} 

In Section \ref{sec:backtoprev}, we apply these approaches to construct confidence intervals for seroprevalence.  We consider several choices of test statistic, including maximum likelihood estimators and generalized likelihood ratio statistics. We demonstrate with simulation that the intervals based on asymptotic or bootstrap approximation have coverage probabilities that, at empirically relevant parameter values and sample sizes, are close to, but potentially below, the nominal level and are approximately equi-tailed.  By contrast, the finite-sample valid construction results in longer intervals on average, but always have coverage probabilities that satisfy the coverage requirement.

We contextualize our analysis with data used to estimate seroprevalence at early stages of the 2019 SARS-Cov-2 pandemic. In particular, as a running example, we measure coverage probabilities and average interval lengths for each of the methods  we consider at sample sizes and parameter values close to the estimates and sample sizes of \cite{bendavid_etal_2020} -- a preprint posted on medRxiv on April 11$^\text{th}$, 2020.\footnote{This preprint has subsequently been published as  \cite{bendavid_published}.} This preprint estimates that the number of coronavirus cases in Santa Clara County, California on April 3 - 4, 2020 was more than fifty times larger than the number of officially diagnosed cases, and as a result, received widespread coverage in the popular and scientific press \citep{kolata2020coronavirus, mallapaty2020antibody}. The methods and design of this study -- including the reported confidence intervals -- were questioned by many researchers \citep{eisen_tibshirani_2020}, prompting the release of a revised draft on April 27$^\text{th}$, 2020, which we refer to as \cite{bendavid_etal_2020b}, that integrated additional data.\footnote{See \cite{gelman2020concerns}, \cite{fithian_2020}, and \cite{bennett2020estimating} for further discussion and analysis of \cite{bendavid_etal_2020b}. In particular, these articles highlight issues and propose alternative approaches for combining data measuring false positive rates from different samples.}  Our analysis highlights statistical challenges in seroprevalence surveys at early stages of the spread of infectious diseases, when disease incidence is low and close to uncertain error rates of new diagnostic technologies. 

This paper contributes to the literatures on inference in seroprevalence surveys \citep{rogan1978estimating, hui1980estimating, walter1988estimation}; see \cite{jewell:2004} for a general introduction to epidemiological statistics.  More broadly, we contribute  to the large literature on test inversion. The classical duality between tests and confidence intervals is discussed in Chapter 3 of \cite{tsh:2005}.  Bootstrap approaches to confidence construction based on estimating nuisance parameters are developed in \cite{efron1981nonparametric}, \cite{diciccio1990bootstrap}, and \cite{carpenter1999test}. Conservative approaches to confidence interval construction that maximize $p$-values over an appropriate nuisance parameter space are considered in \cite{bergerboos:1994} and \cite{silvapulle:1996}.  For the problem considered in this paper, \cite{toulis2020estimation} uses test inversion based on a particular choice of test statistic, though the resulting confidence interval is based on projection. \cite{cai2020exact} is more closely related to one of the approaches we consider, and we discuss some important differences in Section \ref{sec:backtoprev}.\footnote{We became aware of \cite{cai2020exact}, which was posted on arXiv on November 29$\text{th}$, 2020, late in the preparation of this paper, on March 3$\text{rd}$, 2021.} \cite{gelman2020bayesian} take a Bayesian approach to the problem studied in this paper, and give a complementary analysis of uncertainty quantification in \cite{bendavid_etal_2020, bendavid_etal_2020b}. 

%%%%%%%%%%%%%%%%%%%%%%%%%%%%%%%%%%%%%%%%%%%%%%%%%%%%%%%%%%%
\section{Standard Interval Constructions in Seroprevalence Surveys\label{sec: problem}}
%%%%%%%%%%%%%%%%%%%%%%%%%%%%%%%%%%%%%%%%%%%%%%%%%%%%%%%%%%%

A standard seroprevalence survey entails the collection of antibody test results from three samples of individuals of sizes $n_1$, $n_2$, and $n_3$, with $n = (n_1 , n_2 , n_3 )^{\top}$. The first sample is selected at random from the population under study. All individuals in the second sample have not had the disease of interest and all individuals in the third sample have had the disease of interest.\footnote{For example, in \cite{bendavid_etal_2020}, the second sample was composed of blood samples taken before the COVID-19 epidemic and the third sample was composed of blood samples taken from patients who had recovered from confirmed cases of COVID-19.} We let $X = (X_1 , X_2 , X_3 )^\top$ denote the number of positive antibody test results in the corresponding samples. It is assumed that each $X_i$ has a binomial distribution with success probability $p_i$ and is independent of the other samples.\footnote{We assume the sample sizes are small relative to population size so that the difference between sampling with and without replacement is negligible.} The quantities $1-p_2$ and $p_3$ are referred to as the specificity and sensitivity of the test, respectively. We assume that the test has diagnostic value in the sense that $p_2<p_3$, and so $p_1$ necessarily satisfies $p_2 \leq p_1 \leq p_3$.\footnote{Note that some of the methods developed in Section \ref{sec:backtoprev} will not require $p_2<p_3$.} Thus, the parameter $p=\left(p_1,p_2,p_3\right)^\top$ exists in the parameter space
\[
\Omega=\left\{ p \in \left[0,1\right]^{3}:p_{2}\leq p_{1}\leq p_{3}, p_2 < p_3 \right\} .
\]

We consider confidence intervals for the probability $\pi$ that an individual randomly selected from the population under study has had the disease. By the law of total of probability, $p_1 = p_2\left(1-\pi\right) + p_3\pi,$ and so
\begin{equation}\label{equation:sero}
\pi= \pi (p) = \left(p_{1}-p_{2}\right)/\left(p_{3}-p_{2}\right)~.
\end{equation}
We refer to $\pi$ as seroprevalence. A natural estimate of $\pi$ is given by  $\check{\pi}_n = \pi ( \check{p}_n )$, where $\check{p}_{n}=\left(\check{p}_{n,1},\check{p}_{n,2},\check{p}_{n,3}\right)^{\top}$ and $\check{p}_{n,i}=X_i/n_i$ is the usual empirical frequency for group $i$. We let the maximum likelihood estimator (MLE) of $p$ for the model $p \in \Omega$ be denoted by $\hat p_n$, where $\hat{p}_{n}=\left(\hat{p}_{n,1},\hat{p}_{n,2},\hat{p}_{n,3}\right)^{\top}$.\footnote{Typically, $\check p_n$ and $\hat p_n$ agree, with the exception occurring if $\check p_n \notin \Omega$.} Accordingly, the MLE of $\pi$ for the model $p \in \Omega$ is given by $\hat \pi_n = \pi ( \hat p_n )$.

The most obvious approach to constructing confidence intervals for $\pi$ is to approximate the finite-sample distribution of $\hat{\pi}_n$ with its limiting normal distribution. The variance of the normal limiting distribution of $\hat \pi_n$ is given by
\begin{equation}\label{equation:Vp}
{V}_{\hat {\pi}_n} ( p ) =
\frac{1}{\left( p_3 - p_2  \right)^2} \sigma_1^2 \left( p_1 \right)+
\frac{\left({p}_{1} - {p}_{3}\right)^2}{\left({p}_{3} - {p}_{2}\right)^4} \sigma_2^2\left({p}_{2}\right) +
\frac{\left({p}_{2}-{p}_{1}\right)^2}{\left({p}_{3}-{p}_{2}\right)^4} \sigma_3^2\left({p}_{3}\right),
\end{equation}
where
$
\sigma_i^2\left({p}_{i}\right) = p_i \left(1-{p}_{i}\right)/n_i.
$
This leads to the standard Wald or delta method confidence  interval
%\begin{equation}\label{equation:CIwald}
%CI_{\hat{\pi}_n,\Delta} =
$\left(\hat{\pi}_n \pm z_{1 - \alpha/2}\sqrt{{V}_{\hat{\pi}_n}\left(\hat{p}_n\right)}\right)$,
%\end{equation}
where $z_{1 - \alpha/2}$ is the $1-\alpha/2$ quantile of the standard normal cumulative distribution function $\Phi ( \cdot )$.  This construction was used in  \cite{bendavid_etal_2020}.

As outlined in the introduction, the Wald interval may perform poorly in finite-samples due to discreteness of the data or the proximity of parameter values to the boundaries of their spaces. To address some of these issues, \cite{bendavid_etal_2020b} apply the percentile bootstrap confidence interval developed in \cite{efron1981nonparametric}. A refinement of this interval construction, called the $BC_a$ interval \citep{efron1987better}, is also applicable to this problem, with bias and acceleration constants estimated with the formula given in \cite{efron1987better} and \cite{diciccio1995bootstrap}.

The Wald and bootstrap intervals  are approximate. In contrast, it may be desirable to construct intervals that ensure coverage of at least $1-\alpha$ in finite samples, particularly if there are concerns that the finite-sample distribution of $\hat{\pi}_n$ is not well-approximated by a normal distribution. A  simple, but crude, approach to constructing finite-sample valid confidence intervals is projection. In particular, suppose that $R_{1- \alpha}$ is a joint confidence region for $p$ of nominal level $1- \alpha$. The projection method simply constructs the confidence interval
$
I_{1- \alpha } = \{\pi(p):  p \in  R_{1- \alpha}\}.
$
The chance that $\pi \in I_{1- \alpha}$ is bounded below by the chance that $p \in R_{1- \alpha}$. Thus, if $R_{1- \alpha}$ has a guaranteed coverage of $1- \alpha$, then so does $I_{1- \alpha}$. For example, one possible choice of joint confidence region is the rectangle $ R_{1- \alpha } =\prod_{i=1}^3  I_{i, 1- \gamma},$ where $I_{j, 1- \gamma}$ is a  nominal $1- \gamma$ confidence interval for $p_j$ and $\gamma$ is taken to satisfy $(1- \gamma)^3 = 1- \alpha$.\footnote{In our implementation, we apply the standard \cite{clopper1934use} confidence intervals for $p_j$, as they have guaranteed coverage in finite-samples. Other choices exist, however, and in particular the intervals recommended in \cite{brown2001interval} may perform well. Alternatively, the region $R_{1- \alpha }$ can be constructed by inverting likelihood ratio tests, but would incur a significantly larger computational cost. A related approach is developed in \cite{toulis2020estimation}.}  In this case, the computational cost of the projection interval $I_{1- \alpha}$ is minimal, as $\pi (p)$ is monotone increasing in $p_1$ and monotone decreasing in each of $p_2$ and $p_3$ as $p$ varies on the parameter space $\Omega$. Projection intervals are easy to implement, but are generally wide and conservative, in that the true coverage is often larger than the nominal level.

To assess the finite-sample performance of the delta method, bootstrap, and projection confidence intervals, we estimate their coverage probabilities and average lengths at parameterizations close to the sample size and estimates of \cite{bendavid_etal_2020}. In this study, $n_1 = 3300$ participants were recruited for serologic testing for SARS-CoV-2 antibodies. The total number of positive tests was $X_1 = 50$. The authors use $n_2 = 401$ pre-COVID era blood samples to measure the specificity of their test, of which only $X_2 = 2$ samples tested positive. Similarly, the authors use $n_3 = 122$ blood samples from confirmed COVID-19 patients, of which $X_3 = 103$ samples tested positive.\footnote{The specificity and sensitivity samples combine data provided by the test manufacturer and additional tests run at the Stanford. We refer the reader to the statistical appendices of \cite{bendavid_etal_2020,bendavid_etal_2020b} for further details. In \cite{bendavid_etal_2020b} it was revealed that there was an error in the recording of the sensitivity sample, i.e., that there were two fewer positive tests than reported. We adhere to the data as reported in \cite{bendavid_etal_2020}.} The realization of the MLE of $\pi$ for these data is $\hat{\pi}_n = 0.012$.\footnote{ \cite{bendavid_etal_2020} report an alternative estimate of seroprevalence in which the demographics of their sample are weighted to match the overall demographics of Santa Clara County. We briefly discuss the application of the general methods developed in Section \ref{sec: inversion intervals} to this setting in Section \ref{sec:conclusion}, and view further consideration as  a useful extension. \cite{gelman2020bayesian} give a Bayesian approach that accommodates sample weights. In contemporaneous work, \cite{cai2020exact} also address the case where are samples are reweighted according to population characteristics.} Table \ref{table: distribution} reports the nominal 95\% confidence intervals constructed with the standard approaches discussed above. 

\begin{table}

\begin{centering}
\small
\begin{tabular}{ccccc}
\toprule 
 & Interval & Ave. Length & Ave. Length vs. Delta Method & Coverage\tabularnewline
\midrule
\midrule 
Delta Method & [0.003,0.022] & 0.0185 & 1.000 & 0.904\tabularnewline
\midrule 
Percentile Bootstrap & [0.001,0.021] & 0.0186 & 1.005 & 0.895\tabularnewline
\midrule 
$BC_{\alpha}$  Bootstrap & [0.001,0.020] & 0.0191 & 1.028 & 0.895\tabularnewline
\midrule 
Projection & [0.001,0.028] & 0.0270 & 1.4578 & 1.000\tabularnewline
\bottomrule
\end{tabular}
\par\end{centering}
\caption{\label{table: distribution}Average Interval Length and Coverage of Nominal 95\% Confidence Intervals for $\pi$}
{\footnotesize{}Notes: Table \ref{table: distribution} reports the delta method, percentile bootstrap, $BC_{\alpha}$ bootstrap, and projection confidence intervals, at nominal level 95\% computed on data from \cite{bendavid_etal_2020}. Estimates of the average length and coverage for these intervals at sample sizes $n$ and estimated values $\hat{p}_{n}$ from this study are also displayed. Estimates of average length and coverage are taken over 100,000 bootstrap replicates of $X$ at the sample size $n$ and the estimated parameters $\hat{p}_{n}$ from this study.}{\footnotesize\par}

\end{table}

For each parameter (e.g., $p_1$), we simulate replicates of $X$ at each value of a grid around the estimated value of the parameter, holding the other five parameters fixed at their estimated values (e.g., $\hat{p}_{n,2}$, $\hat{p}_{n,3}$, $n_1$, $n_2$, $n_3$). For each method at each combination of parameter values and sample sizes, we compute the proportion of replicates for which the true value of $\pi$ (i.e., the value of $\pi$ associated with the parameterization) is below, contained in, or above the corresponding confidence interval with nominal coverage probability $\alpha=0.05$.

\begin{figure}[t]
\begin{centering}
\caption{\label{fig: distribution} Coverage Performance for Standard Interval Constructions}
\begin{tabular}{c}
\includegraphics[scale=0.6]{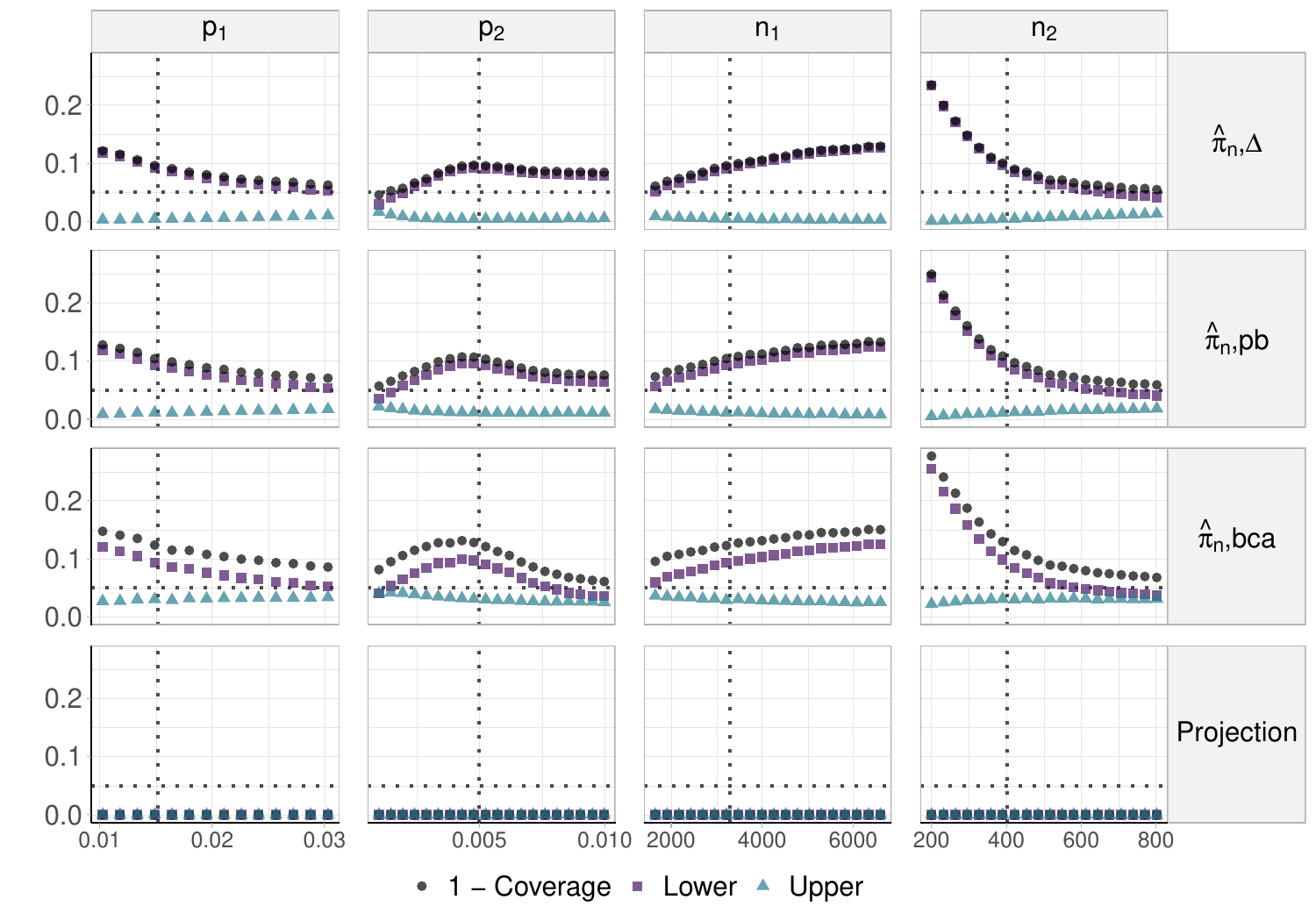}\\
\end{tabular}
\end{centering}
{\footnotesize{}Notes: Figure \ref{fig: distribution} displays estimates of the coverage probabilities of the delta method $\left(\hat{\pi}_{n},\Delta\right)$, percentile bootstrap $\left(\hat{\pi}_n,\text{pb}\right)$, $BC_{\alpha}$ bootstrap $\left(\hat{\pi}_n,\text{bca}\right)$, and projection intervals at parameter values close to the estimate $\hat{p}_{n}$ and sample size $n$ of \cite{bendavid_etal_2020} as specified in Section \ref{sec: problem}. The nominal coverage probability is $0.95$ and is denoted by the horizontal dotted line. The black dots denote one minus the proportion of replicates for which the true value of $\pi$ falls in the realized confidence intervals, i.e., one minus the estimated coverage probability. The purple squares and blue  triangles denote the proportion of replicates that fall below and above realized confidence intervals, respectively. The vertical dotted line denotes the estimated value of $\hat{p}_{n,1}$, $\hat{p}_{n,2}$ or sample size  $n_1$,  $n_2$ for \cite{bendavid_etal_2020}. }{\footnotesize\par}
\end{figure}

Figure \ref{fig: distribution} displays the results of this Monte Carlo experiment for each interval construction at parameter values around $\hat{p}_{n,1}$, $\hat{p}_{n,2}$, $n_1,$ and $n_2$.\footnote{There is little variation in the coverage probabilities for parameter values around $p_3$ and $n_3$, so we omit the results of this experiment for the sake of clarity.} The black dots display one minus the proportion of replicates for which the realized confidence interval contains the true value of $\pi$, i.e., one minus the estimated coverage of the confidence interval. Additionally, Table \ref{table: distribution} reports estimates of the coverage and average length of each interval taken over 100,000 bootstrap replicates at the sample sizes $n$ and estimated values $\hat{p}_{n}$ from \cite{bendavid_etal_2020}.

We find that the delta method and bootstrap intervals are quite liberal. In most cases the estimated coverage of a nominal $95\%$ interval is below $90\%$. The estimated coverage decreases sharply as $n_2$ and $p_1$ become small and is not equi-tailed, in the sense that the proportions of replicates that fall below and above the confidence intervals are not approximately equal. By contrast, the projection method intervals are quite conservative. They are approximately 45\% longer than the delta method intervals at sample sizes $n$ and estimated values $\hat{p}_{n}$ from \cite{bendavid_etal_2020}. These findings motivate the development of approximate and finite-sample valid alternative methods for constructing confidence intervals that have less erratic coverage probabilities. 

%%%%%%%%%%%%%%%%%%%%%%%%%%%%%%%%%%%%%%%%%%%%%%%%%%%%%%%%%%%
\section{Test Inversion \label{sec: inversion intervals}}
%%%%%%%%%%%%%%%%%%%%%%%%%%%%%%%%%%%%%%%%%%%%%%%%%%%%%%%%%%%
In this section, we consider both approximate and finite-sample valid approaches to the general problem of constructing test-inversion confidence intervals for parameters $\theta$, where the corresponding null hypothesis $H\left(\theta_0\right): \theta = \theta_0$ is composite. 
To this end, we require a more general notation. Suppose data $X$ follows a general parametric model indexed by a parameter $( \theta , \vartheta )$ in parameter space $\bar \Omega$. The parameter of interest $\theta$ is real-valued, and the nuisance parameter $\vartheta$ is finite dimensional. For a fixed value $\theta_0$, the parameter space for the nuisance parameter $\vartheta$ is denoted by
\[
\bar \Omega ( \theta_0 ) = \{  (\theta , \vartheta ) \in \bar \Omega: \theta =  \theta_0 \}~.
\]
Observe that for the case of seroprevalence $\pi$, we have that $\theta = \pi$ and can assign $\vartheta = (p_1 , p_3)$.

Test inversion reduces the problem of confidence interval construction for $\theta$ to the problem of testing $H( \theta_0 ) : \theta = \theta_0$ against $\theta > \theta_0$ and $\theta < \theta_0$. Consider a test of the null hypothesis $H\left(\theta_0\right): \theta = \theta_0$ against the alternative $\theta > \theta_0$. A $1-\alpha/2$ lower confidence bound for $\theta$ constructed with test inversion is given by the infimum of the set of $\theta_0$ such that $H\left(\theta_0\right)$ is not rejected at level $\alpha/2$ against the alternative $\theta>\theta_0$, which we denote by $L_{1-\alpha/2}$. A $1-\alpha/2$ upper bound for $\theta$, $U_{1- \alpha/2}$ may be constructed analogously by testing $H\left(\theta_0\right)$ against the alternative $\theta<\theta_0$. Thus, a $1-\alpha$ confidence interval is given by $\left[L_{1 - \alpha/2},U_{1 - \alpha/2}\right]$. 

%%%%%%%%%%%%%%%%%%%%%%%%%%%%%%%%%%
\subsection{Simple Null Hypotheses \label{sec: simple}}
%%%%%%%%%%%%%%%%%%%%%%%%%%%%%%%%%%

It is illustrative to assume that the nuisance parameter $\vartheta = \vartheta_0 $ is known. In this case, the null hypothesis  $H( \theta_0 )$ is simple and one-sided tests that control the level at $\alpha /2$ are easily constructed. Consider the test that rejects $H( \theta_0 )$ against  $\theta > \theta_0$ for large values of the test statistic $T_n =   T_n (X)$. The cumulative distribution function of $T_n$ is given by
$ F_{n, \theta , \vartheta} (t ) = { \mathbb P}_{\theta , \vartheta} \{ T_n  \le t \}.$
Additionally, define the related quantity $F_{n, \theta , \vartheta} (t^- ) = { \mathbb P}_{\theta , \vartheta} \{ T_n  < t \},$ and let $t_0$ denote the observed value of $T_n$. The null probability that $T_n  \ge t_0$ is given by
\begin{equation}\label{equation:pL0}
\hat q_{L, \theta_0 , \vartheta_0 } = 1- F_{n, \theta_0 , \vartheta_0 } (t_0^-)~,
\end{equation}
and is a valid $p$-value in the sense that the test that rejects when this quantity is $ \le \alpha /2$ has size $\le \alpha /2$; see, e.g.,  Lemma 3.3.1 in \cite{tsh:2005}.\footnote{Throughout, we will denote various $p$-values by $\hat q$ rather than $\hat p$ because $\hat p$ is reserved for various estimates of binomial parameters.}
Thus, a $1- \alpha /2$ confidence set for $\theta$ includes all $\theta_0$ such that $F_{n, \theta_0 , \vartheta_0 } ( t_0^- ) < 1- \alpha /2$. If $F_{n, \theta_0 , \vartheta_0 }  (t_0^- )$ is continuous and strictly monotone decreasing in $\theta_0$, then a $1- \alpha/2$ lower confidence bound $ \theta_L$ may be obtained by solving 
\begin{equation}\label{equation:thetaL}
F_{ n, \theta_L , \vartheta_0  } ( t_0^-) = 1- \alpha/2~.
\end{equation}
Similarly, an upper confidence bound $\hat \theta_U$ may be obtained by solving $F_{n, \theta_U , \vartheta_0 } ( t_0 ) = \alpha/2$.\footnote{Even in the case that the distribution of $T_n$ is discrete, the function  $F_{\theta_0 , \vartheta_0 }  (t_0^- )$ is typically continuous in $\theta_0$ (as in the binomial case). If not, one could use the infimum over $\theta_0$ such that $F_{\theta_0 , \vartheta_0 }  (t_0^- )  < 1 - \alpha/2 $ as a lower bound. Note that, in general, may wish to test at each endpoint of a reported confidence interval to determine whether it should be a closed or open interval. We simply take the conservative approach and report closed intervals.} Thus, $[  \theta_L , \theta_U ]$ is a $1- \alpha$ confidence interval for $\theta$. For a single binomial parameter, this construction gives the classical \cite{clopper1934use} interval.
 
 %%%%%%%%%%%%%%%%%%%%%%%%%%%%%%%%%%
\subsection{An Approximate Approach \label{sec: approximate}}
%%%%%%%%%%%%%%%%%%%%%%%%%%%%%%%%%%
 
If the nuisance parameter $\vartheta$ is unknown, then it may be approximated. In particular, if $\hat \vartheta ( \theta_0 )$ is the MLE for $\vartheta$ subject to the constraint $\theta = \theta_0$, then the infeasible $p$-value (\ref{equation:pL0}) can be replaced with
\begin{equation}\label{equation:pL1}
\hat q_{L, \theta_0 ,  \hat \vartheta ( \theta_0  ) } = 1- F_{n, \theta_0 , \hat \vartheta ( \theta_0 ) } (t_0^-)~,
\end{equation}
where $F_{n, \theta_0 , \hat \vartheta ( \theta_0 )}$ is approximated either analytically or with the parametric bootstrap.\footnote{An alternative, related, approach proceeds by imposing and integrating out a potentially uninformative prior on the nuisance parameter, and then constructs test statistics from the resultant pseudo-likelihood function (see e.g., \cite{severini1999relationship} and \cite{datta2012probability}).} Accordingly, the infeasible confidence interval $[ \theta_L, \theta_U ]$ is replaced with the feasible confidence interval $ [ \hat \theta_L , \hat \theta_U ]$, where, the endpoints $\hat \theta_L$ and $\hat \theta_U $ are the values of $\theta_0$ that satisfy
\begin{equation}\label{equation:thetahatL}
F_{n, \hat \theta_L ,  \hat \vartheta ( \theta_0 )    } ( t_0^- ) = 1- \alpha/2~
\quad\text{and}\quad 
F_{n, \hat \theta_U, \hat \vartheta ( \theta_0 ) } ( t_0 ) = \alpha/2~,
\end{equation}
respectively. In other words, either Wald or parametric bootstrap tests are constructed for each $\theta_0$, where the distribution of the test statistic $T_n$ is determined under the parameter $( \theta_0 , \hat \vartheta ( \theta_0 ) )$. This approach was used in  \cite{diciccio1990bootstrap} and \cite{diciccio1995bootstrap}.   

In this approximate approach, the family of distributions indexed by $(\theta ,  \vartheta )$ has been reduced to an approximate least favorable one-dimensional family of distributions governed by the parameter $( \theta _0 , \hat \vartheta ( \theta_0 ) )$ as $\theta_0$ varies. This approach implicitly orthogonalizes the parameter of interest with respect to the nuisance parameter, so that the effect of estimating the nuisance parameter is negligible to second-order and then typically results in second-order accurate confidence intervals.  See \cite{cox:reid:1987} for a discussion of the role of orthogonal parameterizations for inference about a scalar parameter in the presence of nuisance parameters.

%%%%%%%%%%%%%%%%%%%%%%%%%%%%%%%%%%
\subsection{An Infeasible Finite-Sample Approach}\label{section:grid}
%%%%%%%%%%%%%%%%%%%%%%%%%%%%%%%%%%

The quality of the coverage probability of the approximate intervals considered in the previous section will depend on the quality of the approximation of $\hat \vartheta ( \theta_0 )$ to the true value of the nuisance parameter $\vartheta_0$ and the quality of the analytic or bootstrap approximation to the finite-sample distribution $F_{n,\theta_0,\hat \vartheta ( \theta_0 )}$. In situations in which qualities of these approximations are in doubt, e.g., due to discreteness or the proximity of true parameters to the boundary of their parameter spaces, intervals that ensure coverage of at least $1- \alpha$ in finite samples may be desirable.  An infeasible approach to constructing such intervals proceeds by taking the supremum of the $p$-values over all possible values of the nuisance component $\vartheta$, giving
\begin{equation}\label{equation:dec1}
\hat q_{L, \theta_0 , \sup} = 
\sup_{ (\theta_0 , \vartheta )  \in \bar \Omega ( \theta_0 )  } \left(1- F_{n, \theta_0 , \vartheta} ( t_0^- )\right)~,
\end{equation}
with finite-sample validity following from
\[
{\mathbb P}_{\theta_0 , \vartheta_0 }  \{  \hat q_{L, \theta_0 , \sup}  \le u \} 
 \le  {\mathbb P}_{\theta_0 , \vartheta_0 } \{ 1- F_{n, \theta_0 , \vartheta_0 } ( T_n^-  ) \le u \} \le u.
\]
Note that $p$-values of this form may be conservative, as the supremum over $\vartheta$ may be obtained at a value far from $\vartheta_0$.\footnote{On the other hand, if the distribution of the test statistic does not vary much with $\vartheta$, then these $p$-values will not be overly conservative. Therefore, it pays to choose a test statistic that is nearly pivotal, in the sense that its distribution does not depend heavily on $\vartheta$.} To address this issue, one can restrict the space of values for $\vartheta$ that are considered by first constructing a $1- \gamma$ confidence region for $\vartheta$. Such an approach is considered in \cite{bergerboos:1994}, \cite{silvapulle:1996}, and  \cite{rsw:2014}. This refined approach proceeds as follows. Fix a small number $\gamma$ and let  $I_{1- \gamma}$ be a $1- \gamma $ confidence region for $\vartheta$.\footnote{The difficulty of constructing such a region depends on the specific model. In the sequel, we focus attention on cases in which confidence
regions $I_{1-\gamma}$ can be formed by taking the cartesian product of
exact marginal intervals for each component of $\vartheta$.} Consider the modified $p$-value defined by
\begin{equation}\label{equation:dec2L}
\hat q_{L, \theta_0 , I_{1- \gamma} } = 
\begin{cases}
\sup_{ \vartheta  \in I_{1- \gamma}  } 
\left(1- F_{n, \theta_0 , \vartheta} ( t_0^- )\right) + \gamma &  
\text{if } \{\vartheta \in I_{1- \gamma} : (\theta_0, \vartheta) \in \bar{\Omega}(\theta_0)\} \neq \emptyset \\ 
\gamma  &  \text{otherwise}~.
\end{cases}
\end{equation}
The $p$-value obtained by (\ref{equation:dec2L}) is valid in finite samples; see   \cite{bergerboos:1994} or \cite{silvapulle:1996}. Thus, the one-sided test that rejects when $\hat q_{L, \theta_0 , I_{1- \gamma} }\le \alpha/2$ leads to a $1- \alpha/2$ lower confidence bound for $\theta$.  

%%%%%%%%%%%%%%%%%%%%%%%%%%%%%%%%%%
\subsection{A Feasible Finite-Sample Approach}\label{section:restrictgrid}
%%%%%%%%%%%%%%%%%%%%%%%%%%%%%%%%%%

The finite-sample valid approach considered in the previous section is infeasible, as it involves computing a supremum over an infinite set $\bar \Omega ( \theta_0 )$. A natural approximation to this approach is to approximate $\bar \Omega ( \theta_0 )$ with a finite discretization, so that the supremum is replaced by maximum over a finite set of values on a grid.  In particular, if $G( \theta_0 )$ denotes a finite grid over the space $\bar \Omega ( \theta_0 )$, then (\ref{equation:dec1}) can be approximated by
\begin{equation}\label{equation:dec1prime}
\hat q_{L, \theta_0 , \max } =  \max_{ (\theta_0 , \vartheta )  \in G ( \theta_0 )  } 1- F_{n, \theta_0 , \vartheta} ( t_0^- )~.
\end{equation}
Similarly, the refinement given in (\ref{equation:dec2L}) can be approximated by replacing $I_{1-\gamma}$ with $\hat I_{1-\gamma}$, where  $\hat I_{1- \gamma}$ denotes a finite grid (or $\epsilon$-net) approximating $I_{1-\gamma}$.\footnote{In contemporaneous work, \cite{cai2020exact} use this construction to develop confidence intervals for seroprevalence that ensure finite-sample Type 1 error control up to the error induced by the finiteness of $G( \theta_0 )$.}

We develop a modification to this construction that provably maintains finite-sample Type 1 error control for testing $H( \theta_0 )$ by directly accounting for the approximation error induced by a finite discretization of  $\bar \Omega ( \theta_0 )$. Towards this end, we require additional structure. Suppose now that the components of the data  $X = ( X_1 , \ldots , X_k )^{\top}$ are independent, that the distribution of $X_i$ depends on a parameter $\beta_i$, with $\beta = ( \beta_ 1, \ldots , \beta_k )\in\Omega$, and that the family of distributions for $X_i$ has a monotone likelihood ratio in $X_i$. As before, interest focuses on a real-valued parameter $\theta = f ( \beta_1 , \ldots , \beta_k )$, with the nuisance parameter $\vartheta$ given by $\beta_{-1}=(\beta_2,\ldots,\beta_k)$. That is, we assume that the model can be equivalently parameterized by $(\theta,\vartheta)\in\bar{\Omega}$ or through $\beta\in\Omega$, i.e., that the mapping from $\beta$ to $(\theta,\vartheta)$ is one-to-one. For a fixed value $\theta_0$, the parameter space for $\beta$ is given by $\Omega(\theta_0) = \left\{\beta \in \Omega :  f(\beta) = \theta_0 \right\}.$

As before, let $T_n = T_n ( X_1 , \ldots , X_k )$ be a test statistic for testing $H( \theta_0 )$, with $t_0$ denoting its realized value. Assume that $T_n$ is monotone with respect to each each component $X_i$.\footnote{If $T_n$ is monotone decreasing with respect to a particular component, say $X_j$, then $X_j$ can be replaced by $- X_j$, whose family of distributions is then  monotone increasing with respect to $- \beta_j$.} Let $J_{n, \beta} ( \cdot )$ denote the cumulative distribution function of $T_n$ for the $\beta$- parametrization, so that $J_{n, \beta} (t) =  {\mathbb P}_{\beta} \{ T_n \le t \},$ and let $\hat \beta ( \theta_0 )$ denote the MLE for $\beta$ subject to the constraint that $\theta = \theta_0$.  In this case, for example, we can represent the approximate $p$-value defined in Section \ref{sec: approximate} with 
$
\hat{q}_{L,\theta_0, \hat{\vartheta}(\theta_0)} = 1 - J_{n, \hat \beta ( \theta_0 ) } (t_0^-),
$
and similarly for the other $p$-values previously introduced.

We replace the supremum over $I_{1-\gamma}$ in (\ref{equation:dec2L}) with a finite maximum while maintaining Type 1  control. Consider a partition of the values of $\vartheta$ in  $I_{1-\gamma}$ into $r$ regions  $E_1 , \ldots , E_r$. In our implementation, each region is given by a hyperrectangle of the form
$\prod_{i=2}^k[ \beta_i',\beta_i'']$, though this is not essential. For each region $E_j$,  let $\underline{\beta_{-1}}(j) = (\underline{\beta_2} (j)  , \ldots , \underline{\beta_k} (j))$ be the vector giving the smallest value that all but the first component of $\beta$ takes on in $E_j$; that is
$
\underline{\beta_i} (j) = \inf \{ \beta_i : \beta_{-1} \in E_j \}.
$
Analogously, let $\bar{\beta}_{-1}(j)$ be the vector giving the largest value that all but the first component of $\beta$ takes on in $E_j$. For a hyperrectangle $E_j$, clearly $\underline{\beta_i} (j)  = \beta_i'$ and $\bar{\beta}_i  (j)  = \beta_i''$. Congruently, let 
\begin{equation}\label{equation:beta1_under}
\underline{\beta_1} (j) = \inf \{ \beta_1 : \beta \in \Omega(\theta_0), \beta_{-1} \in E_j \}
\quad\text{and}\quad
\bar{\beta}_1 (j) = \sup \{ \beta_1 :  \beta \in \Omega(\theta_0), \beta_{-1} \in E_j \}
\end{equation}
denote the smallest and largest values that the first component of $\beta$ takes on $\Omega(\theta_0)$ for $\beta_{-1}$ in $E_j$. If the infimum or supremum in (\ref{equation:beta1_under}) is over a non-empty set, then define $s_L(j) = J_{n, \bar{\beta} (j) } (t^-)$ and $s_U(j) = J_{n, \underline{\beta} (j) } (t)$, where
$\bar{\beta} (j) = ( \bar{\beta}_1 (j) , \bar{\beta}_{-1} (j))$ and $\underline{\beta} (j) = ( \underline{\beta_1} (j), \underline{\beta_{-1}} (j))$. If there is no $\beta$ in $\Omega(\theta_0)$ with $( \beta_2 , \ldots , \beta_k )$ in $E_j$, then set $s_L(j) = 1$ and $s_U(j)= 0$, respectively.

We construct the $p$-values  
\begin{equation}\label{equation:qtildeL}
\tilde q_{L, \theta_0 , I_{1- \gamma  }} = \max_{1\leq j\leq r} (1- s_L(j)) + \gamma
\quad\text{and}\quad
\tilde q_{U, \theta_0 , I_{1- \gamma  }} = \max_{1\leq j\leq r} s_U(j) + \gamma~
\end{equation}
by taking the maximum over the adjusted $p$-values $1- s_L(j)$ and $s_U(j)$ . This refinement is feasible and valid in finite samples.
\begin{theorem}\label{theorem:2}
Assume that the components of the data  $X = ( X_1 , \ldots , X_k )^{\top}$ are independent, that each component $X_i$ has distribution in a family having a monotone likelihood ratio, and that the statistic $T_n = T_n (X_1 , \ldots , X_k )$ is monotone increasing with respect to each component $X_i$. Let $I_{1- \gamma}$ be a finite-sample valid $1- \gamma$ confidence region for $(\beta_2 , \ldots , \beta_k )$. Then, the $p$-values $\tilde q_{L, \theta_0 , I_{1- \gamma  }} $ and $\tilde q_{U, \theta_0 , I_{1- \gamma  }}$ are valid for testing $H( \theta_0 )$  in the sense that, for any $0 \le u \le 1$ and any $\vartheta$, 
\[
{\mathbb P}_{\theta_0 , \vartheta} \{ \tilde q_{L, \theta_0 , I_{1- \gamma}} \le u \} \le u
\quad\text{and}\quad
{\mathbb P}_{\theta_0 , \vartheta} \{ \tilde q_{U, \theta_0 , I_{1- \gamma}} \le u \} \le u~.
\]
\end{theorem}

\noindent{\sc Proof of Theorem \ref{theorem:2}.}  First, note  that $I_{1- \gamma }$ could be the whole space  $\bar \Omega ( \theta_0 )$ by taking $\gamma = 0$. It follows from  Lemma A.1 in \cite{rsw:2011} (which is a simple generalization of Lemma 3.4.2 in \cite{tsh:2005}) that the family of distributions of $T_n$ satisfies $J_{n, \beta' } (t) \le J_{n, \beta} (t)$ for any $t$,  $\beta = ( \beta_1 , \ldots , \beta_k )$, and $\beta' = ( \beta_1' , \ldots , \beta_k' )$ with $\beta_i' \ge \beta_i $ for  all $i \ge 1$. The same is true if $t$ is replaced by $t^-$. Thus, we have that
$
J_{n, \bar{\beta} (j) } (t^-) \le J_{n, \beta} (t^-)
$
and 
$
J_{n, \underline{\beta} (j) } (t) \ge J_{n, \beta} (t)
$
for any $\beta$ with $\theta = f ( \beta ) = \theta_0$ and  $( \beta_2 , \ldots , \beta_k ) \in E_j$. Therefore, the $p$-value defined by (\ref{equation:dec2L}) satisfies 
\begin{align*}
\hat q_{L, \theta_0 , I_{1- \gamma} }
& \le \max_{1\leq j\leq r} (1- s_L(j)) + \gamma  = \tilde q_{L, \theta_0 , I_{1- \gamma}}~,
\end{align*}
and similarly $\hat q_{U, \theta_0 , I_{1- \gamma} } \leq \tilde q_{U, \theta_0 , I_{1- \gamma}}$ for $\hat q_{U, \theta_0 , I_{1- \gamma} }$ is defined analogously to (\ref{equation:dec2L}) for tests of $H(\theta_0)$ against the alternative $\theta<\theta_0$. Since $\hat q_{L, \theta_0 , I_{1- \gamma }}$ and $\hat q_{U, \theta_0 , I_{1- \gamma }}$ are valid $p$-values, then so are any random variables that are stochastically larger. \hfill \qed

\noindent
Thus, tests based on the $p$-values $\tilde q_{L , \theta_0 , I_{1- \gamma}}$ and  $\tilde q_{U , \theta_0 , I_{1- \gamma}}$ may be used to test $H( \theta_0 )$ and, through test inversion, yield finite-sample valid confidence bounds for $\theta$.  

%%%%%%%%%%%%%%%%%%%%%%%%%%%%%%%%%%%%%%%%%%%%%%%%%%%%%%%%%%%
\section{Test-Inversion Inference for Seroprevalence}\label{sec:backtoprev}
%%%%%%%%%%%%%%%%%%%%%%%%%%%%%%%%%%%%%%%%%%%%%%%%%%%%%%%%%%%

In this section, we apply the general methods considered in Section \ref{sec: inversion intervals} to the problem of constructing approximate and finite-sample valid confidence intervals for seroprevalence.

%%%%%%%%%%%%%%%%%%%%%%%%%%%%%%%%%%
\subsection{Test Statistics \label{sec: test stat}}
%%%%%%%%%%%%%%%%%%%%%%%%%%%%%%%%%%

We begin by exhibiting a set of test statistics $T_n$ applicable to our problem. Let $\hat p_n ( \pi_0 )$ denote the MLE for $p$ restricted to $\Omega ( \pi_0 ) = \{ p \in \Omega :  \pi = \pi_0 \}$.\footnote{Note that there is no explicit representation for $\hat p_n ( \pi_0 )$, but we may compute its value by solving a convex program.} A natural choice for the test statistic $T_n$ is the difference between $\hat{\pi}_n$ and $\pi_0$. This statistic can be Studentized with an estimate of its standard deviation, giving
\[
\tilde{\pi}_n\left(\pi_0\right)=\left(\hat{\pi}_n-\pi_0\right) \Big{/} \sqrt{{V}_{\hat{\pi}_n}\left(\hat{p}_{n}\right)}~,
\]
where $V_{ \hat{\pi}_n} (  p )$ is given in (\ref{equation:Vp}). Alternatively, it can be Studentized with an estimate of its standard deviation  under the constraint $\pi=\pi_0$, giving the test statistic
\[
\tilde{\pi}_{n,C}\left(\pi_0\right) = \left(\hat{\pi}_n-\pi_0\right) \Big{/} \sqrt{{V}_{\hat{\pi}_n}\left(\hat{p}_{n} ( \pi_0 ) \right)}.
\]
Imposing the restriction $\pi=\pi_0$ explicitly in the estimate of the variance may provide a more accurate approximation to the variance of $\hat{\pi}_{n}$ under the null hypothesis. An analogous improvement has been established for the binomial case in \cite{hall1982improving}. 

Observe that as 
$
p_1=p_2\left(1-\pi\right)+p_3\pi,
$
we can rewrite the condition $\pi_0 = \pi$ as the linear restriction $b\left(\pi_0\right)^\top p=0$ where 
$
b\left(\pi_0\right)=\left(1,-\left(1-\pi_0\right),-\pi_0\right)^{\top}.
$
This observation suggests consideration of the linear test statistic 
$
\hat{\phi}_n\left(\pi_0\right)=b(\pi_0)^\top \hat{p}_n.
$
Moreover, as the variance of $\hat{\phi}_n\left(\pi_0\right)$ is exactly equal to 
\begin{equation}
\label{eq: var phi}
V_{\hat{\phi}_n\left(\pi_0\right)}\left(p\right) = 
\sigma_1^2\left(p_1\right) + \left(1-\pi_0\right)^2\sigma_2^2\left(p_2\right) + \pi_0^2\sigma_3^2\left(p_3\right),
\end{equation}
and can be estimated with the plug-in estimator $ V_{\hat{\phi}_n\left(\pi_0\right)}\left(\hat{p}_n\right)$, the test statistic $\hat{\phi}_n\left(\pi_0\right)$ can be Studentized, giving the alternative test statistic
\[
\tilde{\phi}_n\left(\pi_0\right)= \hat{\phi}_n\left(\pi_0\right) \big{/} \sqrt{{V}_{\hat{\phi}\left(\pi_0\right)}\left(\hat{p}_n\right)}.
\]
In turn, we can Studentize $\hat{\phi}_n\left(\pi_0\right)$ with an estimate of its variance under the restriction $\pi_0 = \pi$, giving the statistic
\[
\tilde{\phi}_{n,C}\left(\pi_0\right)= \hat{\phi}_n\left(\pi_0\right) \big{/} \sqrt{{V}_{\hat{\phi}\left(\pi_0\right)}\left(\hat{p}_{n} ( \pi_0 ) \right)}.
\] 
Observe that $\hat{\phi}_n\left(\pi_0\right)$ is well-defined if $p_2 = p_3$, and so $\hat{\phi}_n\left(\pi_0\right)$,  $\tilde{\phi}_n\left(\pi_0\right)$, or $\tilde{\phi}_{n.C}\left(\pi_0\right)$ may be desirable choices in situations where $p_2$ is close to $p_3$.

Alternatively, we can use statistics based on the likelihood function. In particular, let $L\left(p \mid x\right)$ be the likelihood function, given by 
$
L\left(p \mid x\right) = \prod_{1\leq i \leq 3}{n_i \choose x_i}p_i^{x_i} \left(1-p_j\right)^{n_i - x_i}.
$
The generalized likelihood ratio test statistic for testing $H( \pi_0 ):~\pi=\pi_{0}$ is given by 
\begin{equation}
W_{n} = W_{n}\left(\pi_{0}\right) 
=2\cdot\sum_{1\leq j \leq3}\left(X_j \log\left(\frac{\hat{p}_{n,j}}{\hat{p}_{n,j} ( \pi_0 ) }\right)+
                                          \left(n_j-X_j\right)\log\left(\frac{1-\hat{p}_{n,j}}{1-\hat{p}_{n,j} ( \pi_0 ) }\right)\right).
\end{equation}
Large values of $W_{n}$ give evidence for both $\pi<\pi_{0}$ and $\pi>\pi_{0}$. To address this issue, we also consider the signed square root likelihood ratio statistic for the restriction $\pi=\pi_{0}$, given by 
\[
R_n = R_n \left(\pi_0 \right) = \text{sign}\left(\hat{\pi}_{n}-\pi_{0}\right)\cdot\sqrt{W_{n}\left(\pi_0\right)}~.
\]
Corrections to improve the accuracy of $W_n$ based on its signed square root $R_n$ have a long history; see  \cite{lawley:1956},  \cite{barndorff:1986}, \cite{fraser:reid:1988}, \cite{jensen:1986}, \cite{jensen:1992}, \cite{diciccio:2001} and \cite{leeyoung:2005}.  \cite{frydenberg:jensen:1989} consider the effect of discreteness on the efficacy of corrections to improve asymptotic approximations to the distribution of the likelihood ratio statistic. The statistic $R_n$ can be re-centered and Studentized as
\[
\tilde{R}_n = \left(R_n - m_n^{R}\left(\hat{p}_{n} ( \pi_0 ) \right)\right) \big{/} \sqrt{  V_n^R\left(\hat{p}_{n} ( \pi_0 ) \right) }~,
\]
where $m_{n}^{R}\left(p\right)$ and $V_{n}^{R}\left(p\right)$ denote the mean and variance of $R_{n}$ under $p$, and in practice are computed with the bootstrap under $\hat{p}_{n} ( \pi_0 ) $.

%%%%%%%%%%%%%%%%%%%%%%%%%%%%%%%%%%
\subsection{Approximate Intervals \label{sec: approximate application}}
%%%%%%%%%%%%%%%%%%%%%%%%%%%%%%%%%%

We now outline the application of the approximate intervals developed in Section \ref{sec: approximate} to constructing confidence intervals for seroprevalence with the test statistics formulated in Section \ref{sec: test stat}, and measure their performance in the Monte Carlo experiment developed in Section \ref{sec: problem}. Suppose that we are using the test statistic $T_n$ with observed value $t_0$. 
Let
$
J_{n,p} ^T (t) = \mathbb{P}_p\left \{ T_n\leq t\right \}
$
denote the distribution of the general statistic $T_n$ under $p$, and also introduce the related quantity
$
J_{n,p} ^T (t^-) = \mathbb{P}_p\left \{ T_n< t\right \},
$
which will be of use in computing $p$-values for tests of the null hypothesis $\pi = \pi_0$ against alternatives of the form $\pi > \pi_0$.
The approximate test-inversion intervals are constructed by first computing, for each $\pi_0$, the $p$-values
\[
\hat q_{L, \pi_0 ,  \hat{p}_{n} ( \pi_0 ) } = 1- J^T_{n,  \hat{p}_{n} ( \pi_0 )} (t_0^-)
\quad\text{and}\quad 
\hat q_{U, \pi_0 , \hat{p}_{n} ( \pi_0 ) } =  J^T_{n,  \hat{p}_{n} ( \pi_0 )}   (t_0).
\]
The resultant interval with nominal coverage $1-\alpha$ then takes the form 
\[
\left\{\pi_0 : q_{L, \pi_0 ,  \hat{p}_{n} ( \pi_0 ) } \geq \alpha/2 \quad\text{and}\quad q_{U, \pi_0 , \hat{p}_{n} ( \pi_0 )} \geq \alpha/2 \right\}.
\]
We begin by considering asymptotic approximations to the distribution $J^T_{n,  \hat{p}_{n} ( \pi_0 )}$ for different test statistics. 

Observe that, under the null hypothesis $H\left(\pi_0\right)$ and provided that $p$ is not on the boundary of $\Omega$, the test statistics $\tilde{\pi}_{n,C}( \pi_0 )$, $\tilde{\phi}_{n,C}\left(\pi_0\right)$, and $\tilde{R}_n$ are asymptotically $\mathcal{N}\left(0,1\right)$, and $W_n$ is asymptotically $\chi_1^2$. Thus, if we set $T_n$ equal to any of the asymptotically normal statistics, we can approximate $J_{n,  \hat{p}_{n} ( \pi_0 )}$ with a standard normal distribution. Likewise, we may apply a $\chi_1^2$ approximation if we set  $T_n$ equal to $W_n$.\footnote{Observe that using normal approximations to $\hat{\pi}_n$ or $\tilde{\pi}_n$ is equivalent to constructing Wald intervals for these statistics. For that reason, we focus on statistics that make explicit use of the null hypothesis restriction $\pi = \pi_0$.}

Table \ref{table: asymptotic} reports realizations of these approximate confidence for the observed values from \cite{bendavid_etal_2020}. Additionally, Table \ref{table: asymptotic} presents estimates of coverage and average interval length taken over 10,000 bootstrap replicates computed at the $n$ and estimate $\hat{p}_{n}$ from this study.  Notably, each of these intervals now includes zero.\footnote{If the null hypothesis $\pi = 0$ is of particular interest or concern, then there exists an exact uniformly most powerful unbiased level $\alpha$ test for the equivalent problem of testing $p_1 = p_2$ against $p_1 > p_2$. This is a conditional one-sided binomial test; see Section 4.5 of \cite{tsh:2005}. Such a test does not exist for other values of $\pi_0$.} These interval constructions are roughly the same length, on average, as the delta method intervals, but have coverage probability significantly closer to the nominal level. 

\begin{table}
\begin{centering}
\small
\begin{tabular}{ccccc}
\toprule 
Statistic & Interval & Ave. Length & Ave. Length vs. Delta Method & Coverage\tabularnewline
\midrule
\midrule 
$\tilde{\pi}_{n,C}$ & [0.000,0.020] & 0.0193 & 1.0404 & 0.963\tabularnewline
\midrule 
$\tilde{\phi}_{n,C}$ & [0.000,0.020] & 0.0191 & 1.0302 & 0.963\tabularnewline
\midrule 
$W_{n}$ & [0.000,0.021] & 0.0177 & 0.9563 & 0.927\tabularnewline
\midrule 
$\tilde{R}_{n}$ & [0.000,0.021] & 0.0181 & 0.9758 & 0.950\tabularnewline
\bottomrule
\end{tabular}
\par\end{centering}
\caption{\label{table: asymptotic} Average Interval Length and Coverage for Test-Inversion Nominal 95\% Confidence  Intervals  for Seroprevalence Using Asymptotic Approximation}
{\footnotesize{}Notes: Table \ref{table: asymptotic} reports the approximate test-inversion confidence intervals, constructed with an asymptotic approximation to test statistic null distributions, computed on data from \cite{bendavid_etal_2020}. Estimates of the average length and coverage for these intervals at the $n$ and estimate $\hat{p}_{n}$ from this study are also displayed. Estimates of average length and coverage are taken over 10,000 bootstrap replicates of $X$ at the sample size $n$ and the estimated parameters $\hat{p}_{n}$ from this study.}{\footnotesize\par}
\end{table}

\begin{figure}[t]
\begin{centering}
\caption{\label{fig: asymptotic} Coverage Performance for Test-Inversion Intervals Based on Asymptotic Approximation}
\begin{tabular}{c}
\includegraphics[scale=0.6]{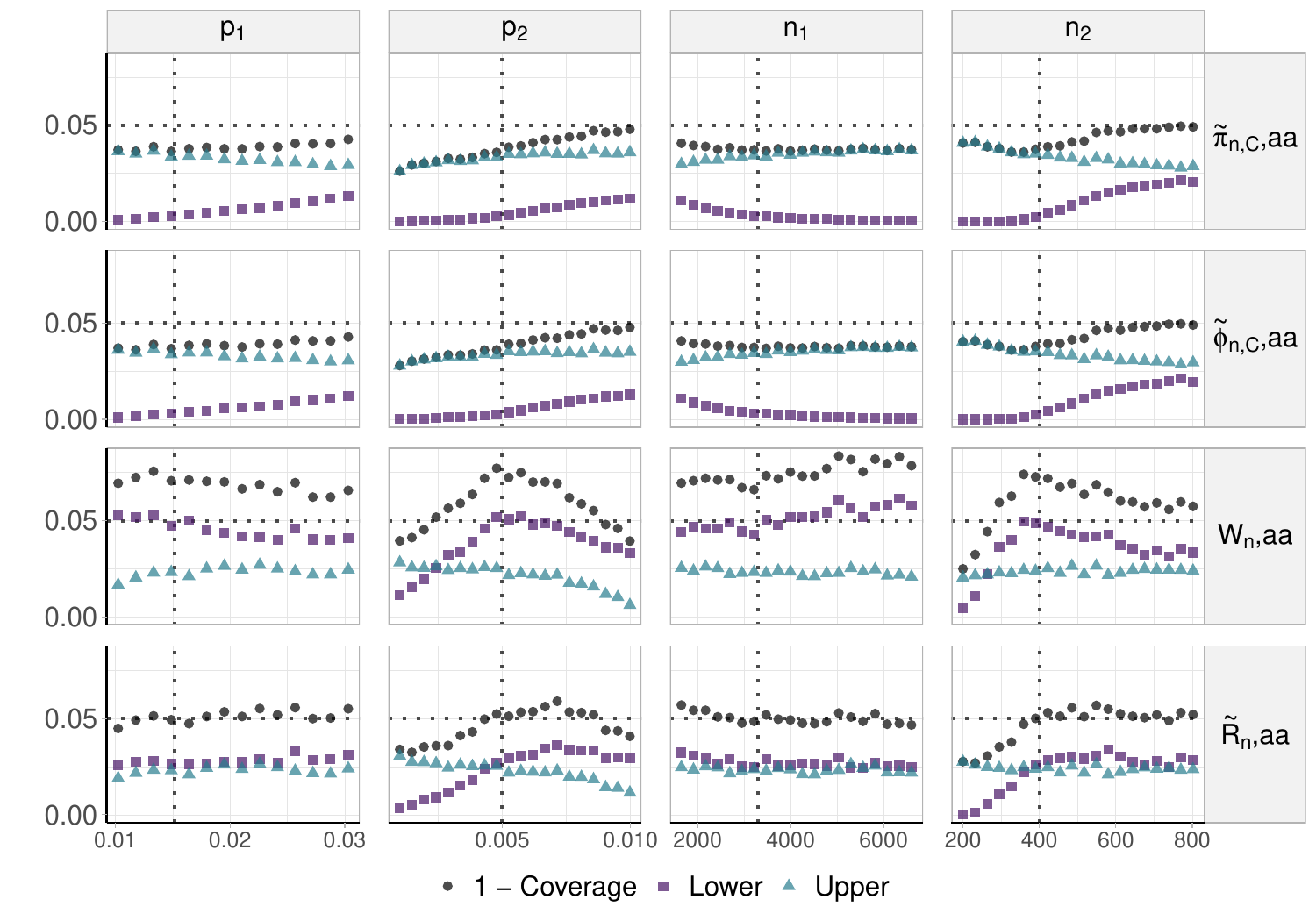}\\
\end{tabular}
\par\end{centering}
{\footnotesize{}Notes: Figure \ref{fig: asymptotic} displays estimates of the coverage probabilities of the approximate confidence intervals constructed with an asymptotic approximation to test statistic null distributions. The nominal coverage probability is $0.95$ and is denoted by the horizontal dotted line. Estimates of the coverage for the interval constructed with the test statistic $T_n$ are denoted by ``$T_n,\text{aa}$'' and are computed at parameter values close to the estimates $\hat{p}_n$ and sample size $n$ of \cite{bendavid_etal_2020} as specified in Section \ref{sec: problem}. The black dots denote one minus the proportion of replicates for which the true value of $\pi$ falls in the realized confidence intervals, i.e., one minus the estimated coverage probability. The purple squares  and blue  triangles denote the proportion of replicates that fall below and above realized confidence intervals, respectively. The vertical dotted line denotes the   estimated value of $\hat{p}_{n,1}$, $\hat{p}_{n,2}$, or sample $n_1$,  $n_2$ for \cite{bendavid_etal_2020}.}{\footnotesize\par}
\end{figure}

Figure \ref{fig: asymptotic} displays estimates of the coverage probabilities for these intervals in the Monte Carlo experiment developed in Section \ref{sec: problem}. Recall that the black dots display one minus the estimates of the coverage probabilities of the respective intervals, and that the purple squares  and blue  triangles display the proportion of the replicates in which the true value of $\pi$ falls below and above the realized confidence interval, respectively. In contrast to the results for the standard methods displayed in Figure \ref{fig: distribution}, we estimate that the coverage probabilities for these methods are very close to the nominal value of $95\%$ for most parameterizations. The intervals constructed with $W_n$ and $\tilde{R}_n$ are the most equi-tailed.

\begin{table}[t]
\begin{centering}
\small
\begin{tabular}{ccccc}
\toprule 
Statistic & Interval & Ave. Length & Ave. Length vs. Delta Method & Coverage\tabularnewline
\midrule
\midrule 
$\hat{\pi}_{n}$ & [0.000,0.021] & 0.0189 & 1.0186 & 0.965\tabularnewline
\midrule 
$\hat{\phi}_{n}$ & [0.000,0.021] & 0.0188 & 1.0133 & 0.962\tabularnewline
\midrule 
$\tilde{\pi}_{n}$ & [0.000,0.021] & 0.0188 & 1.0150 & 0.953\tabularnewline
\midrule 
$\tilde{\phi}_{n}$ & [0.000,0.021] & 0.0188 & 1.0141 & 0.953\tabularnewline
\midrule 
$W_{n}$ & [0.000,0.021] & 0.0181 & 0.9788 & 0.946\tabularnewline
\midrule 
$R_{n}$ & [0.000,0.021] & 0.0186 & 1.0035 & 0.951\tabularnewline
\bottomrule
\end{tabular}
\par\end{centering}
\caption{\label{table: bootstrap} Average Interval Length and Coverage for Approximate Test-Inversion Nominal 95\% Confidence  Intervals Based on the Bootstrap}
{\footnotesize{}Notes: Table \ref{table: bootstrap} reports the approximate test-inversion confidence intervals, constructed with a parametric bootstrap approximation to  null distributions of test statistics, computed on data from \cite{bendavid_etal_2020}. Estimates of the average length and coverage for these intervals at sample size $n$ and estimate $\hat{p}_{n}$ from this study are also displayed. Estimates of average length and coverage are taken over 10,000 bootstrap replicates of $X$ at the sample size $n$ and estimate  $\hat{p}_{n}$ from this study.}{\footnotesize\par}
\end{table}

Next, we refine this approach by directly computing $J_{n,  \hat{p}_{n} ( \pi_0 )}$ with the bootstrap. This method is more accurate, but can be more computationally expensive. In particular, choosing test statistics that make use of the constrained MLE $\hat{p}_{n}(\pi_0)$ requires solving the associated convex program for each bootstrap replicate. As a result, for this case, we focus attention on test statistics that do not make use of $\hat{p}_{n}(\pi_0)$. Table \ref{table: bootstrap} reports realizations of these approximate confidence intervals  computed on data from \cite{bendavid_etal_2020}, in addition to estimates of the coverage and average interval length at the $n$  and estimate $\hat{p}_{n}$ for this study. Again, each of these intervals include zero, are roughly the same length as the delta-method intervals, and have coverage probability close to the nominal level.

\begin{figure}[p]
\begin{centering}
\caption{\label{fig: bootstrap} Coverage Performance for Test-Inversion Intervals Based on Bootstrap Approximation}
\begin{tabular}{c}
\includegraphics[scale=0.6]{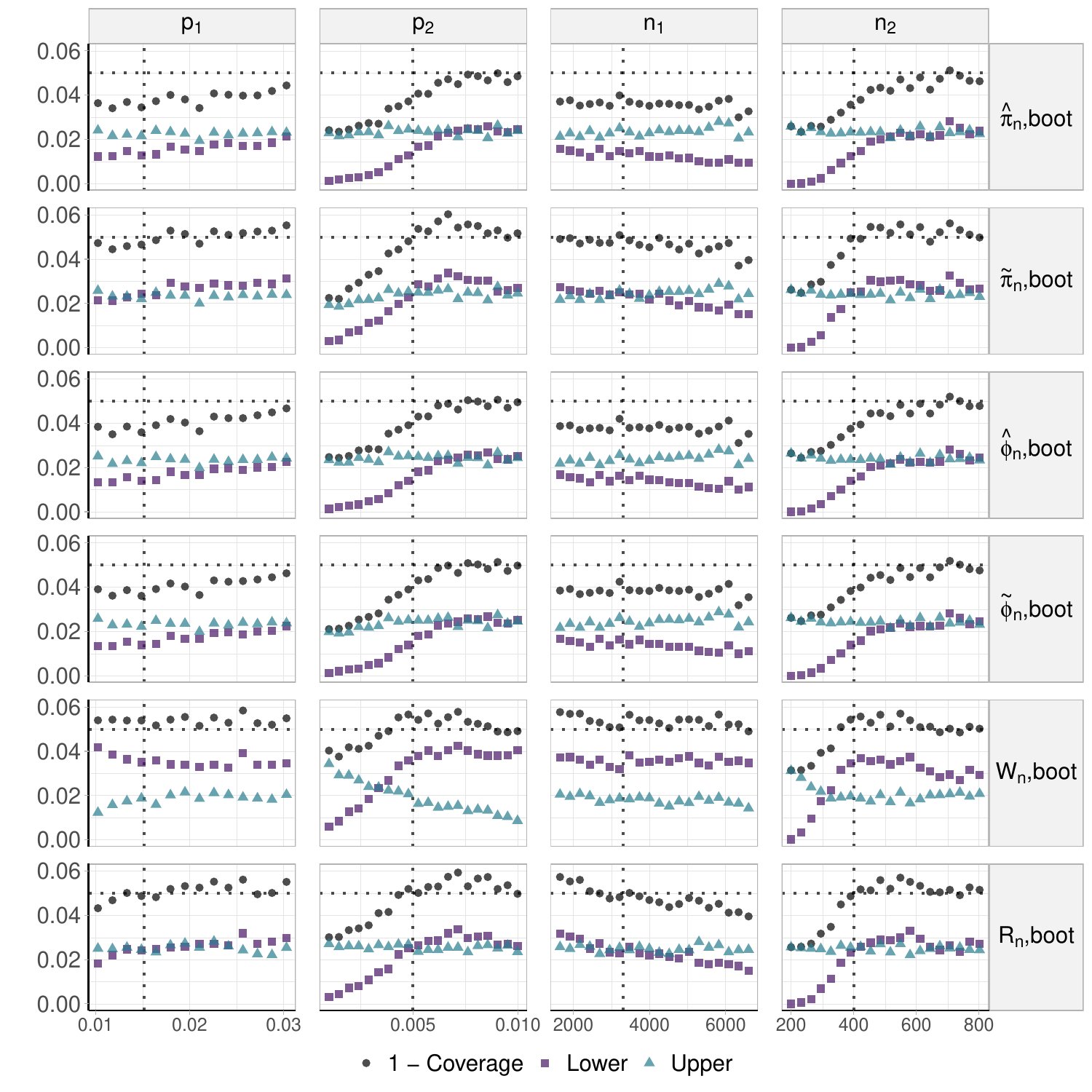}\\
\end{tabular}
\par\end{centering}
{\footnotesize{}Notes: Figure \ref{fig: bootstrap} displays estimates of the coverage probabilities of the approximate confidence intervals constructed with the bootstrap. The nominal coverage probability is $0.95$ and is denoted by the horizontal dotted line. Estimates of the coverage for the interval constructed with the test statistic $T_n$ are denoted by ``$T_n,\text{boot}$'' and are computed at parameter values close to the estimates $\hat{p}_n$ and sample size $n$ of \cite{bendavid_etal_2020} as specified in Section \ref{sec: problem}. The black dots denote one minus the proportion of replicates for which the true value of $\pi$ fall in the realized confidence intervals, i.e., one minus the estimated coverage probability. The purple  squares and blue  triangles denote the proportion of replicates that fall below and above realized confidence intervals, respectively. The vertical dotted line denotes the estimate $\hat{p}_{n,1}$, $\hat{p}_{n,2}$ or sample size $n_1$,  $n_2$ for \cite{bendavid_etal_2020}.}{\footnotesize\par}
\end{figure}

Figure \ref{fig: bootstrap} displays estimates of the coverage probabilities in the same Monte Carlo experiment developed in Section \ref{sec: problem}. Again, these intervals have coverage close to the nominal value and are approximately equi-tailed. The interval constructed with $R_n$ is most equi-tailed and appears to be the least sensitive to perturbations in $p_2$ and $n_2$.

%%%%%%%%%%%%%%%%%%%%%%%%%%%%%%%%%%
\subsection{Finite-Sample Valid Intervals}
%%%%%%%%%%%%%%%%%%%%%%%%%%%%%%%%%%

We now turn to the application of the finite-sample valid intervals discussed in Section \ref{section:restrictgrid}. We focus our development on the test statistic $\hat{\phi}_n(\pi_0)$ as it is linear, and therefore monotone with respect to each sample $X_i$, as is required.\footnote{One may also consider the test statistic $\hat \pi_n$, as $\pi(\cdot)$ is monotone with respect to each component as long as $p \in \Omega$. } 

To begin, we partition the parameter space $\Omega$ into a parameter of interest and a nuisance component. Recall that finite-sample exact intervals formed by maximizing $p$-values over a nuisance space will perform best if the distribution of the chosen test statistic does not vary much with the nuisance parameter. For small values of $\pi_0$, the variance of $\hat{\phi}_n(\pi_0)$ is insensitive to changes in $p_3$, as the variance $\sigma_3^2(p_3)$ enters into (\ref{eq: var phi}) linearly and scaled by $\pi_0^2$. Additionally, for sample sizes comparable to the measurements taken in \cite{bendavid_etal_2020}, where $n_1$  is much larger than $n_2$, the variance of $\hat{\phi}_n(\pi_0)$ will be less sensitive to changes in $p_1$ than to changes in $p_2$. Thus, we set the nuisance component $\vartheta = (p_1,p_3)$, giving the parameterization $(\pi,\vartheta)$.\footnote{We note that for different sample sizes, it may be attractive to set the nuisance component $\vartheta = (p_2,p_3)$. For example, \cite{bendavid_etal_2020b} --  the April 27$^\text{th}$ draft of \cite{bendavid_etal_2020} -- includes larger sensitivity and specificity samples $n_2$ and $n_3$. In particular, the specificity sample $n_2$ was increased from 401 to 3324. These additional data were are aggregated over several samples taken at different times and locations.  \cite{gelman2020concerns}, \cite{fithian_2020}, and \cite{bennett2020estimating} highlight issues with this aggregation.

The choice $\vartheta = (p_2 , p_3)$ also has computation advantages. In particular, by the identity $p_1=p_2\left(1-\pi\right)+p_3\pi$, for any value of seroprevalence $\pi_0$ and any values of $p_2$ and $p_3$ satisfying $p_2 < p_3$, there is a value of $p_1$ that satisfies $p_2 \leq p_1 \leq p_3$ such that $\pi_0 = (p_1 - p_2)/(p_3 - p_2).$ That is, any value of $(p_2,p_3)$ corresponds to a unique value of $p_1$ consistent with a given value of $\pi_0$ and satisfying the a priori restrictions on the parameter space $\Omega$.}

\begin{figure}[t]
\begin{centering}
\caption{\label{fig: initial region} Bootstrap Quantiles and Initial  Nuisance Parameter Confidence Region}
\begin{tabular}{c}
\includegraphics[scale=0.60]{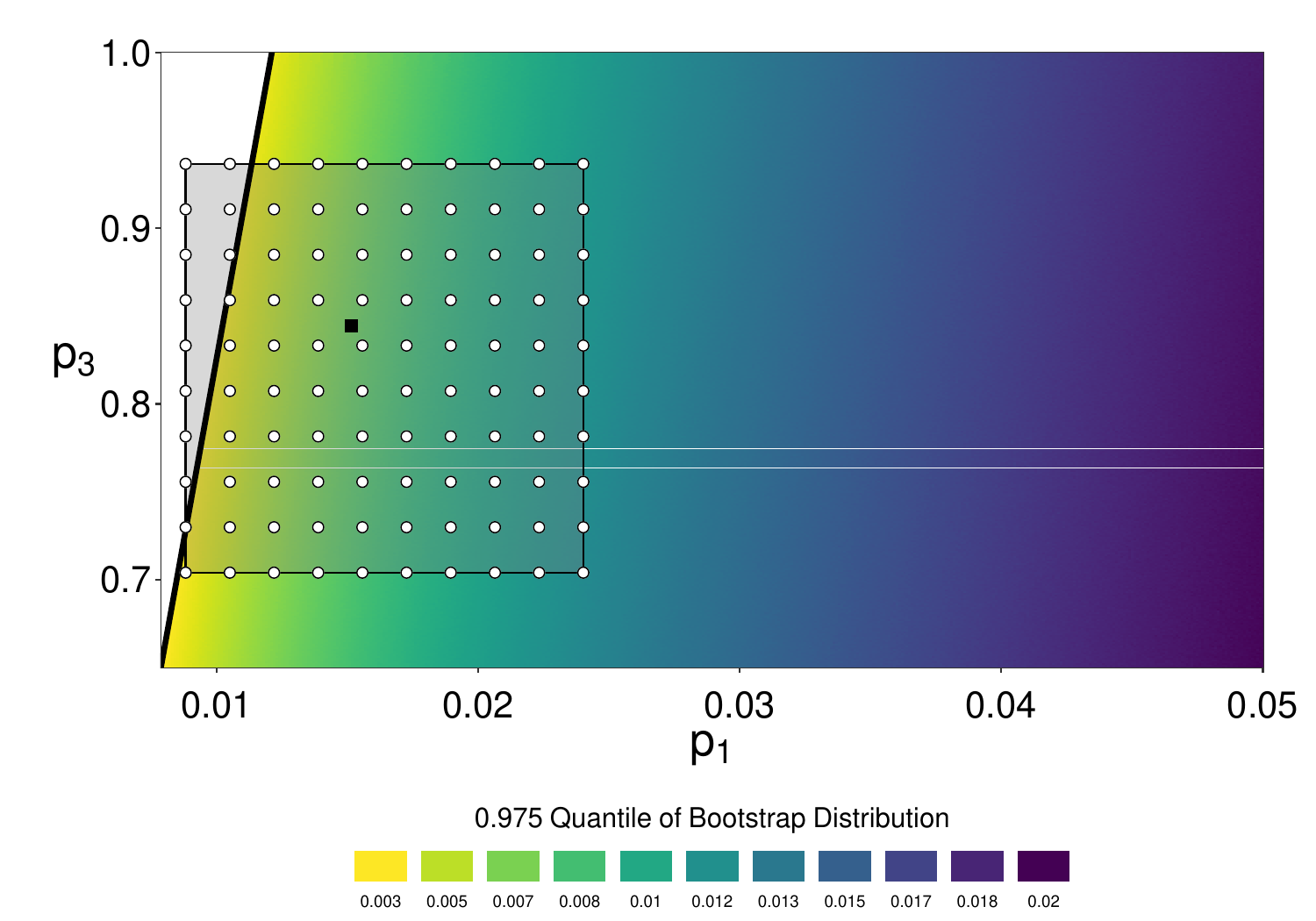}\\
\end{tabular}
\par\end{centering}
{\footnotesize{}Notes: Figure \ref{fig: initial region} displays a heat-map of the $0.975$ quantile of $\hat{\phi}_n(\pi_0)$ under different parameter values $(\pi_0, \vartheta)$, where the sample size and null hypothesis restriction $\pi_0$ equal to the  sample size and estimated prevalence $\hat{\pi}_n$ from \cite{bendavid_etal_2020}. The black line exhibits the boundary of the parameter space $\bar \Omega(\pi_0)$. The black square dot denotes the constrained MLE $\hat{\vartheta}(\pi_0) = (\hat{p}_{n,1} ( \pi_0 ), \hat{p}_{n,3} ( \pi_0))$. With $\gamma = 0.001$, the grey rectangle denotes a $1-\gamma$ confidence region for $\vartheta$ constructed by taking the cartesian product of two $\sqrt{1-\gamma}$ level confidence regions for $p_1$ and $p_3$ each constructed with the method of \cite{clopper1934use}. The white dots denote a $10\times10$ grid over this space.}{\footnotesize\par}
\end{figure}

To fix ideas, consider Figure \ref{fig: initial region}, which displays a heat-map of the $0.975$ quantile of $\hat{\phi}_n(\pi_0)$ under different values of the nuisance parameter $\vartheta$, with the sample size and the null hypothesis restriction $\pi_0$ equal to the  sample size and estimated prevalence $\hat{\pi}_n$ from \cite{bendavid_etal_2020}. The square black dot denotes the constrained MLE, $\hat{\vartheta}(\pi_0) = (\hat{p}_{n,1} ( \pi_0 ), \hat{p}_{n,3} ( \pi_0))$. The black line exhibits the boundary of the parameter space $\bar \Omega(\pi_0)$. 

Recall that in the constructions of approximate intervals considered in Section \ref{sec: approximate application}, a point $\pi_0$ is excluded from a confidence interval with nominal coverage $0.95$ if the observed value of the chosen test statistic $T_n$ exceeds or falls below the $0.975$ or $0.025$ quantiles of the statistic's finite-sample distribution at the constrained MLE, $\hat{p}_{n} ( \pi_0 )$. However, as illustrated in  Figure \ref{fig: initial region}, the $0.975$ quantile of the bootstrap distribution of $\hat{\phi}_n(\pi_0)$ has considerable variation with the nuisance parameter $\vartheta$. As a result, these approximate intervals will not exactly control the coverage probability in finite samples, as the event that $\vartheta$ differs from $\hat{\vartheta}(\pi_0)$ occurs with positive probability.

In turn, comparing the realized value of a test statistic to quantiles of the statistic's finite-sample distribution at every value of the nuisance component $\vartheta$ is both infeasible, as the space of $\vartheta$ is infinite, and impractical, as it would lead to extremely conservative intervals. In fact, we can see that in Figure \ref{fig: initial region}, the $0.975$ quantile of the bootstrap distribution of $\hat{\phi}_n(\pi_0)$ is approximately four times as large at $p_1 = 0.05$ than at  $p_1 =\hat{p}_{n,1} ( \pi_0 )$.

Thus, the finite-sample approach developed in Section \ref{section:restrictgrid} begins by constructing a $1-\gamma$ confidence region for $\vartheta$ and forming a finite grid over this space. The initial confidence region $I_{1-\gamma}$ is illustrated in Figure \ref{fig: initial region} by the greyed rectangle, and a $10\times10$ grid over this space is illustrated by the grid of white dots. The confidence region $I_{1-\gamma}$ is formed by taking the Cartesian product of two $\sqrt{1-\gamma}$ level confidence regions for $p_1$ and $p_3$, each constructed by using the exact intervals of \cite{clopper1934use}. For the purposes of this figure, we set $\gamma = 0.001$.  This grid partitions the values of $\vartheta$ in $I_{1-\gamma}$ into $r = 81$ rectangles, which we enumerate $E_1, \ldots, E_r$.  Define, for $i = 1,3$, the extreme points $\underline{p_i}(j) = \inf\{p_i : (p_1, p_3)^\top \in E_j\}$ and $\overline{p_i} (j) = \sup\{p_i : (p_1, p_3)^\top \in E_j\}$ as well as 
\begin{align}
\label{eq: p_1}
\underline{p_2}(j) &= \inf\{p_2 : p \in \Omega(\pi_0), (p_1, p_3)^\top \in E_j\} \quad\text{and}\\
\overline{p_2}(j) &= \sup\{p_2 : p \in \Omega(\pi_0), (p_1, p_3)^\top \in E_j\}~.\nonumber
\end{align}
As the test statistic $\hat \phi_n ( \pi_0 )$ is monotone increasing in $X_1$ and monotone decreasing in $X_2$ and $X_3$, define ${p_L}(j)  = \left (\overline{p_1} ( j) , \underline{p_2} (j) , \underline{p_3} (j) \right )$ and ${p_U}(j) =  \left ( \underline{p_1} ( j) , \overline{p_2} (j) , \overline{p_3} (j) \right) $
as well as
\[
s_L(j) = J^{\hat{\phi}(\pi_0)}_{n,  {p_L} (j)} (t_0^-)
\quad\text{and}\quad 
s_U(j) = J^{\hat{\phi}(\pi_0)}_{n,  p_U (j)} (t_0)
\] 
where $s_L(j)$ and  $s_L(j)$ are set equal to 1 and 0, respectively,  if  the infimum or supremum in (\ref{eq: p_1}) are taken over the empty set. Thus, by Theorem \ref{theorem:2} we can construct the finite-sample valid $p$-values
\[
\tilde{q}_{L,\pi_0,I_{1-\gamma}}= \max_{1\leq j \leq r} \left(1 -  s_L(j)\right) + \gamma
\quad\text{and}\quad
\tilde{q}_{U,\pi_0,I_{1-\gamma}}= \max_{1\leq j \leq r}\left(  s_U(j)\right) + \gamma
\]
for testing the null hypothesis $\pi = \pi_0$. Hence, the resultant finite-sample valid interval with nominal coverage $1-\alpha$ takes the form 
$
\left\{\pi_0 : \tilde{q}_{L,\pi_0,I_{1-\gamma}} \geq \alpha/2 \quad\text{and}\quad \tilde{q}_{U,\pi_0,I_{1-\gamma}} \geq \alpha/2 \right\}.
$

This approach is closely related to the method developed in \cite{cai2020exact}, though there are some differences. Roughly, \cite{cai2020exact} compute $p$-values for test of the null hypothesis $\pi = \pi_0$ with the parametric bootstrap using the particular choice of test statistic $\tilde{\pi}_n$ at each point of a grid spanning  a confidence region for the nuisance parameter. Their construction begins by constructing a joint confidence region for all three parameters, while our approach proceeds from a smaller initial region for just $p_1$ and $p_3$.  We make an additional correction for a grid approximation to the nuisance space, which allows us to ensure finite-sample validity. Their construction does not guarantee that the resulting intervals are equi-tailed.

\begin{figure}[t]
\begin{centering}
\caption{\label{fig: exact} Coverage Performance for Finite-Sample Valid Test-Inversion and Projection Intervals}
\begin{tabular}{c}
\includegraphics[scale=0.6]{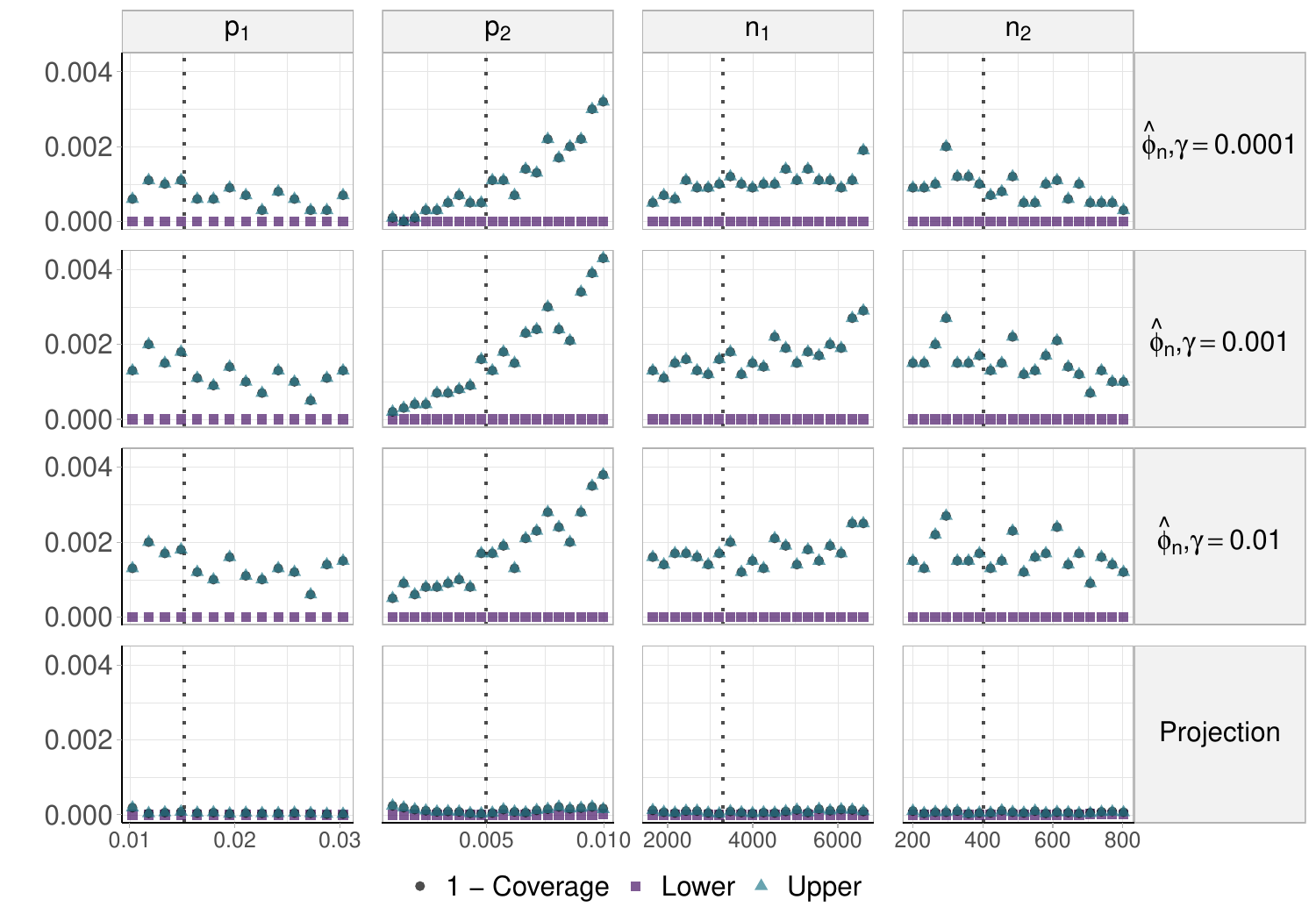}\\
\end{tabular}
\par\end{centering}
{\footnotesize{}Notes: Figure \ref{fig: exact} displays estimates of the coverage probabilities of the finite-sample valid test inversion and projection confidence intervals. The nominal coverage probability is $0.95$. Estimates of the coverage are computed at parameter values close to the estimates $\hat{p}_n$ and sample size $n$ from \cite{bendavid_etal_2020} as specified in Section \ref{sec: problem}. The black dots denote one minus the proportion of replicates for which the true value of $\pi$ falls in the realized confidence intervals, i.e., one minus the estimated coverage probability.  The purple squares  and blue  triangles denote the proportion of replicates that fall below and above realized confidence intervals, respectively.  Note that, in the case of this figure, the black dots and blue triangles are approximately overlaid.The vertical dotted line denotes the estimated value of $\hat{p}_{n,1}$, $\hat{p}_{n,2}$, or sample size $n_1$,  $n_2$ from  \cite{bendavid_etal_2020}.}
\end{figure}

Table \ref{table: exact} reports realizations of these finite-sample valid intervals for several values of $\gamma$, in addition to the projection intervals discussed in Section \ref{sec: problem}, for \cite{bendavid_etal_2020}.\footnote{These results are insensitive to small changes in the grid size $g$.} The table also reports estimates of the coverage and average interval length at the estimated values of $\hat{p}_n$ for this study. The cost of ensuring finite-sample valid coverage is large. The realized intervals are roughly 40\% wider on average than intervals constructed with the delta method, and the coverage is very close to one. Figure \ref{fig: exact} displays estimates of the coverage probabilities for the finite-sample valid intervals as well as the projection intervals in the same Monte Carlo experiment developed in Section \ref{sec: problem}.\footnote{Note that the proportion of Monte Carlo replicates for which the true value of $\pi$ falls below the realized intervals is very close to zero at most parameter values, and so dots denoting one minus the estimated coverage and the proportion of Monte Carlo replicates for which the true value of $\pi$ falls above the realized intervals are approximately overlaid.} Again, the coverage is very close to one at small sample sizes. However, the finite-sample valid intervals outperform the projection intervals. The difference is most salient in the measurements of coverage. 

\begin{table}[t]
\begin{centering}
\small
\begin{tabular}{cccccc}
\toprule 
Method & $\gamma$ & Interval & Ave. Length & Ave. Length vs. Delta Method & Coverage\tabularnewline
\midrule
\midrule 
\multirow{3}{*}{Exact} & 0.0001 & {[}0.000,0.028{]} & 0.0283 & 1.5311 & 0.999\tabularnewline
\cmidrule{2-6} \cmidrule{3-6} \cmidrule{4-6} \cmidrule{5-6} \cmidrule{6-6} 
& 0.0010 & {[}0.000,0.027{]} & 0.0269 & 1.4538 & 0.998\tabularnewline
\cmidrule{2-6} \cmidrule{3-6} \cmidrule{4-6} \cmidrule{5-6} \cmidrule{6-6} 
& 0.0100 & {[}0.000,0.026{]} & 0.0259 & 1.3979 & 0.998\tabularnewline
\cmidrule{1-6} \cmidrule{2-6} \cmidrule{3-6} \cmidrule{4-6} \cmidrule{5-6} \cmidrule{6-6}
Projection &  &  {[}0.001,0.028{]} &  0.0270 & 1.4578 & 1.000 \tabularnewline
\bottomrule
\end{tabular}
\par\end{centering}
\caption{\label{table: exact}  Average Interval Length and Coverage for Finite-Sample Valid Test-Inversion and Projection Nominal 95\% Intervals}
{\footnotesize{}Notes: Table \ref{table: exact} reports the finite-sample valid test-inversion and projection confidence intervals computed on data from \cite{bendavid_etal_2020}. Estimates of the average length and coverage for these intervals at sample size $n$ and estimate $\hat{p}_{n}$ from this study are also displayed. Estimates of average length and coverage are taken over 10,000 bootstrap replicates of $X$ at the sample size $n$ and the estimate $\hat{p}_{n}$ from this study.}{\footnotesize\par}
\end{table}

The additional costs associated with correction of the approximation error induced by a finite discretization of the nuisance space are not overly burdensome. In particular, consider the test inversion intervals for $\pi$ constructed with the $p$-values $\hat q_{L, \pi_0 , \hat{I}_{1-\gamma,g}}$ and $\hat q_{U, \pi_0 , \hat{I}_{1-\gamma,g}} $, where the former $p$-value is defined in (\ref{equation:dec2L}), the latter is defined analogously for upper confidence bounds, and $\hat{I}_{1-\gamma,g}$ is a $g\times g$ grid over the initial confidence region $I_{1-\gamma}$. That is, $\hat{I}_{1-\gamma , g}$ denotes the white dots in Figure \ref{fig: initial region}, where in that case $g=10$. The realized value for these intervals with $g=10$ and $\gamma = 10^{-2}$ for data from  \cite{bendavid_etal_2020} are $[0.000, 0.025]$. For these values of $g$ and $\gamma$, this interval construction has an average length of $0.0249$, which is $34.85$\% longer than the delta-method interval on average, i.e., they are $3.5$\% shorter than the finite-sample valid intervals considered in this section. 

There are several facets of the finite-sample valid confidence intervals considered in this section that could potentially be improved. These include the choice of the nuisance parameter $\vartheta$ that leads to an initial confidence region $I_{1- \gamma}$. Additionally, an initial confidence region may be constructed by taking the product of appropriate one-sided bounds, respectively, rather than using a single joint confidence region for both lower and upper bounds. This change should save roughly $\gamma/2$ in over-coverage. It may be more desirable to use a Studentized test statistic, as its distribution may vary even less within the initial confidence region. However, a nontrivial modification of the correction for the approximation error induced by a finite discretization of the nuisance space is required, since monotonicity may be violated. Lastly, finer grids over the nuisance space may be applied to further reduce the length of intervals.

%%%%%%%%%%%%%%%%%%%%%%%%%%%%%%%%%%%%%%%%%%%%%%%%%%%%%%%%%%%
\section{Conclusion}\label{sec:conclusion}
%%%%%%%%%%%%%%%%%%%%%%%%%%%%%%%%%%%%%%%%%%%%%%%%%%%%%%%%%%%
We demonstrate that standard methods for constructing confidence intervals in basic seroprevalence surveys derived from the delta method, the percentile bootstrap, and the $BC_a$ bootstrap have coverage probabilities that behave erratically and are not consistently near the nominal level at empirically relevant sample sizes and parameter values. By contrast, we show that methods that combine test inversion with the parametric bootstrap lead to stable coverage probabilities that are close to the nominal level across a variety of statistics. Specifically, statistics that are properly Studentized or based on the generalized likelihood ratio statistic exhibit superior performance. Test inversion based on the signed square root generalized likelihood ratio statistic gives the best overall performance in terms of stability and validity over a range of empirically relevant parameterizations and sample sizes. On the other hand, if one desires methods with guaranteed coverage in finite-samples, then we have provided an alternative construction with finite-sample validity, at the cost of coverage above the nominal level and longer intervals on average.

Our reanalysis of the uncertainty in estimates of the seroprevalence of SARS-CoV-2 antibodies in Santa Clara County, California on April 3-4, 2020 suggests that the data collected in \cite{bendavid_etal_2020} are insufficient to rule out small population proportions of SARS-CoV-2 antibodies. However, it is important to note that the maintained assumption -- that the sample of antibody tests administered to the population of interest is collected at random -- is likely not to hold in many applications. In particular, in the case of \cite{bendavid_etal_2020}, it stands to reason that populations that differed in their likelihood of exposure to COVID-19 differed in their likelihood of volunteering for antibody testing. In an attempt to account for these selection effects, \cite{bendavid_etal_2020} apply post-stratification weighting by zip code, sex, and race to match population weights measured with the 2018 American Community Survey. With the use of this re-weighting, estimates of seroprevalence and corresponding confidence intervals appear to increase by approximately one to two percent.\footnote{The demographic data necessary for the replication of this result are not available. \cite{bendavid_etal_2020} report a weighted seroprevalence estimate and confidence interval -- purportedly constructed with the delta method -- of 0.0249 and (0.0201,0.0349), respectively. However, \cite{fithian_tweet} argues convincingly that there were coding errors made in the computation of these intervals. \cite{cai2020exact} report weighted percentile bootstrap confidence intervals of (0.0110,0.0372), where we note that there are small differences in the specificity and sensitivity estimates that they use relative to the data studied in this article.} Consequently, the interpretation of the results of the \cite{bendavid_etal_2020} seroprevalence survey appears to be contingent on the form of population weighting applied. This sensitivity highlights the fundamental importance of high quality data collection in survey design, and supports a view of seroprevalence surveys as an important input into, but not a final answer for, assessments of the progression of early stages of infectious diseases.

The methods discussed in this paper are applicable to, but not tailored for, post-stratification weighted estimation. If there are $S$ strata of the population of interest, with $p_1^i$ denoting the probability that a randomly selected individual in the $i$th stratum tests positive, then seroprevalence in the $i$th stratum is $\pi_i = ( p_1^i - p_2)/(p_3-p_2)$. If the $i$th stratum gets the known weight $w_i$, then the overall seroprevalence is $\pi = \sum_i w_i \pi_i$, which is a function of $S+2$ binomial parameters. However, for even moderately large values of $S$, the finite-sample valid intervals developed in Section \ref{section:restrictgrid} will be computationally expensive due to the need to compute $p$-values for each point in a discretization of an $S+1$ dimensional first-stage confidence region.\footnote{In this case, adapting the finite-sample valid procedures proposed in this article for use with asymptotic approximations to the distribution of the signed square root likelihood ratio statistic (see e.g., \cite{brazzale2007applied, jensen:1986, jensen:1992}) may significantly facilitate computation.} The computational cost of the approximate intervals considered in Section \ref{sec: approximate} will not be higher than for the unweighted problem. We view further consideration of confidence interval constructions that are well suited for post-stratification weighting as a useful direction for further research.\footnote{Both \cite{gelman2020bayesian} and  \cite{cai2020exact} propose approaches to this problem, with the former taking a Bayesian perspective.}

In many applied contexts it is likely valuable -- both for estimation and uncertainty quantification -- to incorporate other forms of information made available in the collection of samples and test characteristics. For example, in \cite{bendavid_etal_2020b} a larger specificity sample is constructed by aggregating several samples from different populations. As indicated in \cite{fithian_2020}, there is evidence that there is greater variation in estimates of specificity across these samples than would be expected if each of the test results in these samples were independent and identically distributed. \cite{gelman2020bayesian} propose a hierarchical approach to accounting for this over-dispersion, suggesting a model in which the specificity parameters of the tests implemented in each sample -- including the sample taken from the population of interest -- are drawn from a pre-specified parametric distribution. As the process generating the specificity samples might tend to be different from the process generating the population of interest (e.g., specificity samples may be drawn from hospital patients local to test manufacturers), we would advocate for an approach in which specificity was modeled as a function of relevant population characteristics. \cite{gelman2020bayesian} also highlight the possibility of incorporating individual-level symptom data; we second this suggestion and view it as a useful direction for further research. 

The methods presented in Section \ref{sec: inversion intervals} apply quite generally to the construction of confidence intervals for real-valued parameters $\theta$. A subclass of problems can be described as follows. Suppose $X_1 , \ldots , X_k$ are independent, with $X_i$ distributed as a binomial with parameters $n_i$ and $p_i$. It is desired to construct a confidence interval for some parameter $\theta = f ( p_1 , \ldots , p_k )$.  In this case, the family of distributions of $X_i$ has monotone likelihood ratio in $X_i$. The application of Theorem \ref{theorem:2} requires specification of a nuisance parameter $\vartheta$, construction of a confidence interval for $\vartheta$, and verification that the chosen test statistic is monotone with respect to each of its components. The latter is straightforward when the test statistic is given by $T_n = f ( \check{p}_1 , \ldots , \check{p}_n )$, where $\check{p}_i = X_i / n_i$. Some important special cases include differences of proportions, measures of relative risk reduction, and odds ratios.\footnote{Uniformly most accurate unbiased confidence bounds exist only for the odds ratio based on classical constructions, but optimality considerations fail for the other parameters; see e.g., Problem 5.29 of \cite{tsh:2005}} \cite{agresti2002unconditional} and \cite{fagerland2015recommended} also consider similar confidence interval constructions for these three parameters that involve minimizing $p$-values over a nuisance parameter space, but do not account for the discretization required.
  
%%%%%%%%%%%%%%%%%%%%%%%%%%%%%%%%%%%%%%%%%%%%%%%%%%%%%%%%%%%
% BIBLIOGRAPHY
%%%%%%%%%%%%%%%%%%%%%%%%%%%%%%%%%%%%%%%%%%%%%%%%%%%%%%%%%%%
\begin{spacing}{1}
\newpage
{\small
{\bibliography{prevalence.bib}}
}
\end{spacing} 
\end{document}